\documentclass[final,3p,times,pdflatex]{elsarticle}
\usepackage{amssymb,bbold}
\usepackage{amsmath}
\usepackage[utf8x]{inputenc}
\usepackage{natbib}
\usepackage{graphicx}
\usepackage{hyperref}
\usepackage{epstopdf}
\usepackage{pifont}
\usepackage{float}

\graphicspath{{pdf/}}

\hypersetup{colorlinks=true,linkcolor=blue,citecolor=blue,filecolor=blue,urlcolor=blue,breaklinks=true}
\biboptions{sort&compress}
\journal{Annals of Physics}
\RequirePackage{color}

\begin{document}

\begin{frontmatter}

\title{Nonlinear refractive index changes and absorption coefficients in mesoscopic ring induced by variable effective mass}

\author[ufes]{Denise Assafrão}
\ead{denise.lima@ufes.br}

\author[ceunes]{A. G. de Lima}
\ead{andre.g.lima@ufes.br}

\author[ufma]{Edilberto O. Silva}
\ead{edilberto.silva@ufma.br}

\author[ufla]{Cleverson Filgueiras}
\ead{cleverson.filgueiras@dfi.ufla.br }

\address[ufes]{Departmento de Física, Universidade Federal do Espírito Santo , Vitória, ES, Brazil}
\address[ceunes]{Departmento de Ciências Naturais, CEUNES, Universidade Federal do Espírito Santo , São Mateus, ES, Brazil}
\address[ufma]{Departamento de F\'{\i}sica, Universidade Federal do Maranh\~{a}o,\\ 65085-580 S\~{a}o Lu\'{\i}s, MA, Brazil}
\address[ufla]{Departamento de F\'{i}sica, Universidade Federal de Lavras,\\ Caixa Postal 3037, 37200-000, Lavras, Minas Gerais, Brazil}

\begin{abstract}
This study explores the linear and nonlinear optical absorption coefficients (OAC) and refractive index changes (RIC) in quantum dot and quantum antidot systems with a position-dependent variable effective mass described as \(m(r)=\alpha r^{\gamma}\). Significant contributions to both linear and nonlinear OAC and RIC are observed. Our findings reveal that variations in \(\gamma\) modify the intersubband dipole matrix elements and energy intervals, leading to noticeable shifts in optical properties. The results show that higher \(\gamma\) values shift resonance peaks towards higher energies, while changes in the oscillator frequency result in abrupt shifts and peak diminutions. These insights provide a deeper understanding of the optical behaviors in the quantum systems under consideration, paving the way for designing devices with optimal efficiency.
\end{abstract}


\end{frontmatter}

\section{\label{sec:Intro}Introduction}

In recent decades, there has been growing interest among researchers in low-dimensional semiconductor structures, driven mainly by their remarkable electronic and optical properties. Nonlinear optics based on two-dimensional (2D) materials represents a significant branch of modern optics \cite{LIU2024393,2d}, emerging as critical components in the construction of solid-state nanodevices \cite{doi:10.1126/science.1261243}. The interconversion between electrical and optical energy is crucial in optoelectronics, where light-emitting devices such as LEDs and diode lasers efficiently convert electrical energy into visible light \cite{bao20192d}. In parallel, photodetectors perform the opposite, converting optical energy into usable electrical signals \cite{WANG2018149}. These energy conversion processes have been explored and refined, driving significant advances in optical communication and light detection technology \cite{LIANG2024104340}.

From a fundamental standpoint, describing optical emission and absorption is essential to these achievements. The optical properties of a two-dimensional quantum ring with a pseudopotential in the presence of an external magnetic field and magnetic flux were theoretically investigated in \cite{LIANG20115818}. Reference \cite{jaouane2022effects} explores the effects of donor impurity in spherical quantum dots with finite confinement potential under an externally applied magnetic field. The eccentric quantum ring under a parallel magnetic field has pronounced effects on two-dimensional GaA's electronic and optical properties \cite{ringparallel}. In the context of the influence of Rashba spin-orbit interaction, it was shown that the geometry of a ring significantly influences the magnitude of the nonlinear absorption coefficient and refractive index changes, allowing for the design of devices with optimal efficiency \cite{rashba}. The influence due to topological defect and Rashba spin-orbit interaction on the optical properties of a two-dimensional GaAs quantum dot is addressed in \cite{rashba}. In \cite{doubleQR}, the optical properties of two-dimensional coupled concentric double quantum rings with a parabolic confinement potential are examined, revealing a noticeable shift in the peak position of the absorption coefficient due to impurities. The optical absorption coefficients of a quantum well strongly depend on its parameters, with values increasing notably when considering electron-phonon interaction \cite{Zhang}. Non-inertial effects can be viewed in \cite{rotating}. The optical absorption of electron gas interacting with various potentials is discussed in Refs. \cite{parabolicpotential,rosenMorse,hellman}.

As well known, the effective mass of electrons in materials has pronounced effects in nonlinear optics \cite{massxpt2}. In Ref. \cite{massxpt}, its impact on the optical properties in coupled quantum dot-rings is explored.

Still, in the context of the above systems, we highlight the models with position-dependent mass (PDM). The study of PDM models is of significant importance in various fields of physics due to their wide range of applications. In quantum mechanics, PDM models provide a more accurate description of particles in inhomogeneous materials where the effective mass varies with position, such as in semiconductor heterostructures and quantum dots \cite{Bastard}. In mesoscopic physics, these models are essential for understanding the electronic properties of systems with varying material composition \cite{Harrison}. Furthermore, in condensed matter physics, PDM models help study excitons and other quasiparticles in complex materials \cite{Morrow}. In atomic physics, PDM models are also applied to describe atoms in varying external fields and environments \cite{vonRoos}.

The effects of spatially varying effective mass distribution on the electronic and optical properties were addressed in \cite{massAnarmonic}, and it was discovered that the optical characteristics of the system could be customized to meet specific needs by adjusting the structural parameters, which dictate size and shape, alongside selecting an appropriate, effective mass distribution. The variability in spatial positioning of mass and confinement potential significantly influences the absorption coefficients and overall changes in refractive index \cite{variablemassDot,massWell,LIMA2023115688}. These diverse applications highlight the relevance of developing and analyzing PDM models in theoretical and applied physics. Inspired by these works, we turn our attention to theoretically investigating the impacts of variable effective electron mass systems on the optical absorption and the refractive index of an electron gas within confined potentials that offer analytical solutions. Quantum systems with variable mass \cite{doi:10.1098/rspa.2022.0200,crevim} have been explored, particularly when investigating problems that are soluble \cite{article1,article2}.

In this work, we explore the impact of position-dependent effective mass on the optical properties of quantum dot and quantum antidot systems. We utilize a radial-dependent mass model to examine how variations in the effective mass distribution, characterized by the power-law dependence \( m(r) = \alpha r^\gamma \), affect the linear and nonlinear optical absorption coefficients (OAC) and refractive index changes (RIC) in these confined quantum systems. The study employs the point canonical transformation (PCT) method, which allows us to derive the energy eigenvalues and wave functions analytically for both quantum dot and quantum antidot potentials. These analytical solutions are then used to calculate the intersubband dipole matrix elements and energy intervals, which directly influence the optical behavior of the systems \cite{Alhaidari2002,Khordad2023,Das2021}. We investigate two distinct cases: quantum dots, characterized by a potential that leads to bound states in a confining environment, and quantum antidots, where the potential allows for extended states \cite{Cruz2022,Saha2022}. For both cases, we study how the parameters \(\gamma\) and \(\omega_0\) modify the optical properties, focusing on the resonance peaks of absorption and the shifts in energy levels. By comparing these cases, we provide insights into the tunability of the optical properties through the control of effective mass distribution \cite{Shi2020}. Our results offer valuable perspectives on how the interplay between spatially varying mass and confinement potentials can be harnessed to optimize the performance of quantum devices, particularly in applications related to optoelectronics and nanotechnology \cite{Mansoor2021,Tao2023}.

This paper is organized as follows: in Section \ref{sec:PCT}, we present a detailed formulation of the quantum dot and quantum antidot models, incorporating a position-dependent effective mass framework. We derive the eigenvalue equations, discuss the confining potentials, and outline the analytical methods employed to obtain the energy levels and wavefunctions. In Section \ref{sec:optics}, we focus on the optical properties of the system, including both linear and nonlinear optical absorption coefficients (OAC) and refractive index changes (RIC). We also provide a comprehensive discussion on how these optical phenomena are influenced by position-dependent mass and external parameters. Section \ref{sec:results} presents and analyzes the numerical results, comparing the effects of different parameters on the optical responses of quantum dots and antidots. Finally, in Section \ref{sec:concl}, we summarize the key findings of the work and highlight the potential applications of these results in the design of optoelectronic devices based on quantum systems with position-dependent mass.

\section{\label{sec:PCT}Radial dependent mass model}

In this section, we present the model that describes the motion of a quantum particle with an effective mass that depends explicitly on position. We shall write the eigenvalue equation and obtain the energies and wave functions, considering the inclusion of a confining radial potential, which will be defined later. We started by writing the general effective Hamiltonian for the case of a specified PDM model, \cite{Alhaidari}
\begin{equation}
   H=\frac{1}{2}\mathbf{P}\frac{1}{m\left(\mathbf{r}\right)}\mathbf{P}+V\left(\mathbf{r}\right)=-\frac{1}{2}\boldsymbol{\nabla}\frac{1}{m\left(\mathbf{r}\right)}\boldsymbol{\nabla}+V\left(\mathbf{r}\right),
   \label{eq:hamiltonian}
\end{equation}
where $m(\mathbf{r})$ and $V(\mathbf{r})$ are real function of the configuration space coordinates, and $\hbar=m_{e}=1$. The radial differential equation derived from the Hamiltonian (\ref{eq:hamiltonian}) for a three-dimensional problem with spherical symmetry gives
\begin{equation}
\left\{ \frac{d^{2}}{dr^{2}}-\frac{l(l+1)}{r^{{2}}}+\frac{m^{\prime}}{m}\left(\frac{1}{r}-\frac{d}{dr}\right)-2m\left[V(r)-E\right]\right\}\phi(r)=0,
\label{eq:variabel_mass}
\end{equation}
where $l$ is the angular momentum quantum number, $m^{\prime}=dm(r)/dr$, and $\Psi(r)=\phi(r)/r$, as usual. On the other hand, we used the three-dimensional isotropic oscillator as a reference \cite{Alhaidari}, defined by
\begin{equation}
\upsilon\left(\rho\right)=\frac{1}{2}\lambda^{4}\rho^{2},
\end{equation}
to solve the time-independent radial Schrödinger wave equation for a constant mass, which reads
\begin{equation}
 \left\{ \frac{d^{2}}{d\rho^{2}}-\frac{\mathcal{L}(\mathcal{L}+1)}{\rho^{2}}-2m\left(\upsilon\left(\rho\right)-\varepsilon\right)\right\} \psi\left(\rho\right)=0, 
 \label{eq:constant_mass}
\end{equation}
where $\mathcal{L}$ is the angular momentum quantum number, and $\varepsilon$ are the energy eigenvalues. In this case, it can be shown that by solving equation \ref{eq:constant_mass}, the energy eigenvalues and energy eigenfunctions are
\begin{equation}
\varepsilon_{n}=\lambda^{2}\left(2n+\mathcal{L}+\frac{3}{2}\right),\label{en}
\end{equation}
and
\begin{equation}
\psi_{n}=a_{n}\left(\lambda \rho\right)^{\mathcal{L}+1} e^{^{-\frac{\lambda^{2}\rho^{2}}{2}}}L_{n}^{\mathcal{L}+2}\left(\lambda^{2}\rho^{2}\right),\label{pn}
\end{equation}
respectively.
By making the change of variable in Eq. (\ref{eq:constant_mass}), $\psi\left(\rho\right)=g(r)\phi(r)$, with $\rho=q(r)$, we obtain the following radial equation:
\begin{equation}
\left\{ \frac{d^{2}}{dr^{2}}+\left(2\frac{\dot{g}}{g}-\frac{\ddot{q}}{q}\right)\frac{d}{dr}+\left(\frac{\ddot{g}}{g}-\frac{\ddot{q}}{\dot{q}}\frac{\dot{g}}{g}\right)- \mathcal{L}\left(\mathcal{L}+1\right)\left(\frac{\dot{q}}{q}\right)^{2} 
-2\left(\dot{q}\right)^{2}\left(\upsilon\left(q\left(r\right)\right)-\varepsilon\right)\right\} \phi(r)=0.
\label{eq:transformation}
\end{equation}
The comparison between Eqs. (\ref{eq:constant_mass}) and (\ref{eq:transformation}) gives $g(r)=\sqrt{\dot{q}/m}$ and the following correspondence:
\begin{equation}
V\left(r\right)-E+\frac{l\left(l+1\right)}{2mr^{2}}=\frac{\left(\dot{q}\right)^{2}}{m}\left(\upsilon\left(q\right)-\varepsilon\right)+\frac{\mathcal{L}(\mathcal{L}+1)}{2m}\left(\frac{\dot{q}}{q}\right)^{2}+
\frac{\dot{m}}{2m^{2}r}+\frac{1}{4m}\left(F(m)-F(\dot{q})\right).
\label{eq:final_equation}
\end{equation}

\begin{figure}[tbh]
\centering
{
\includegraphics[width=.4\textwidth]{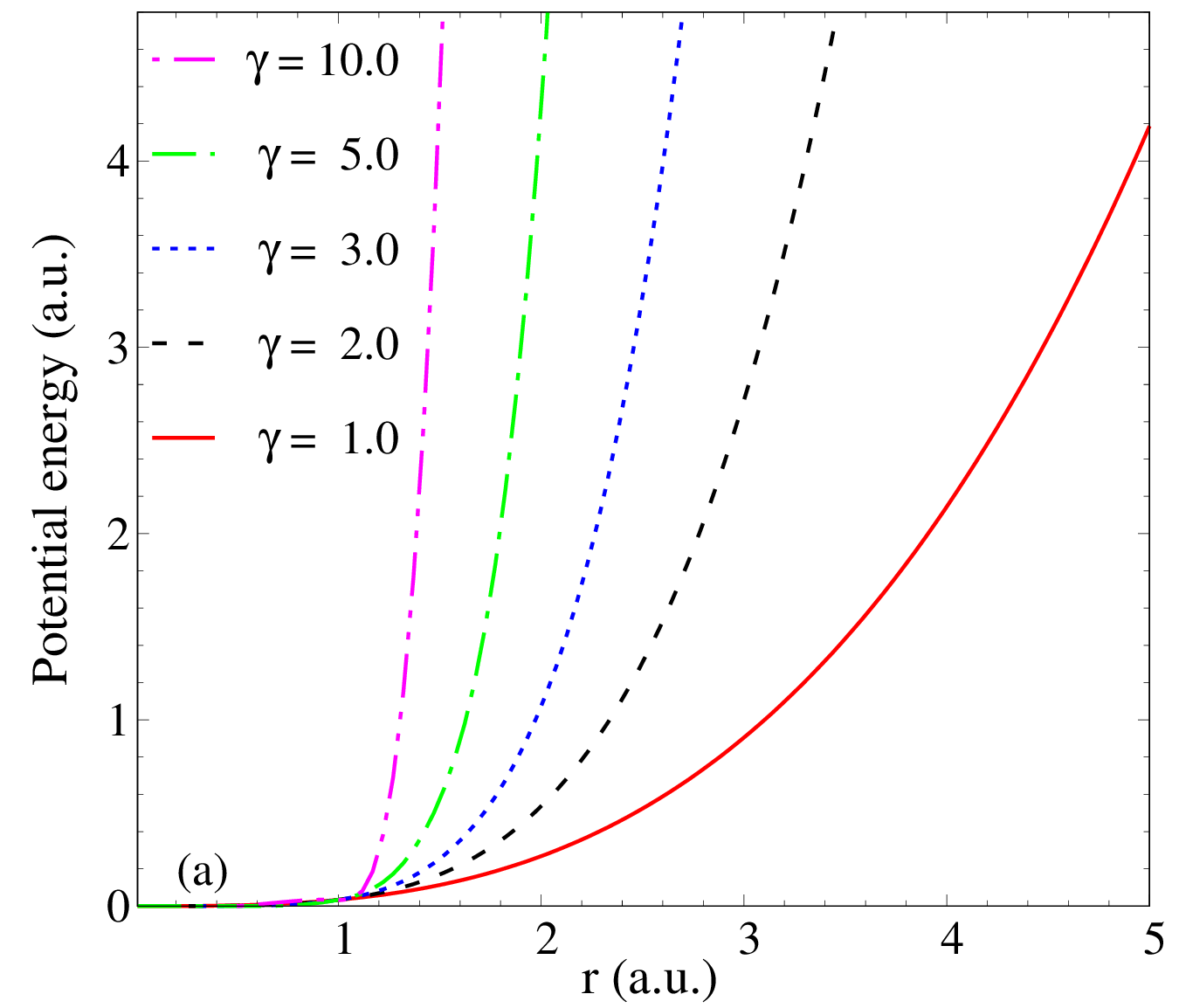}
} 
\quad 
{
\includegraphics[width=.4\textwidth]{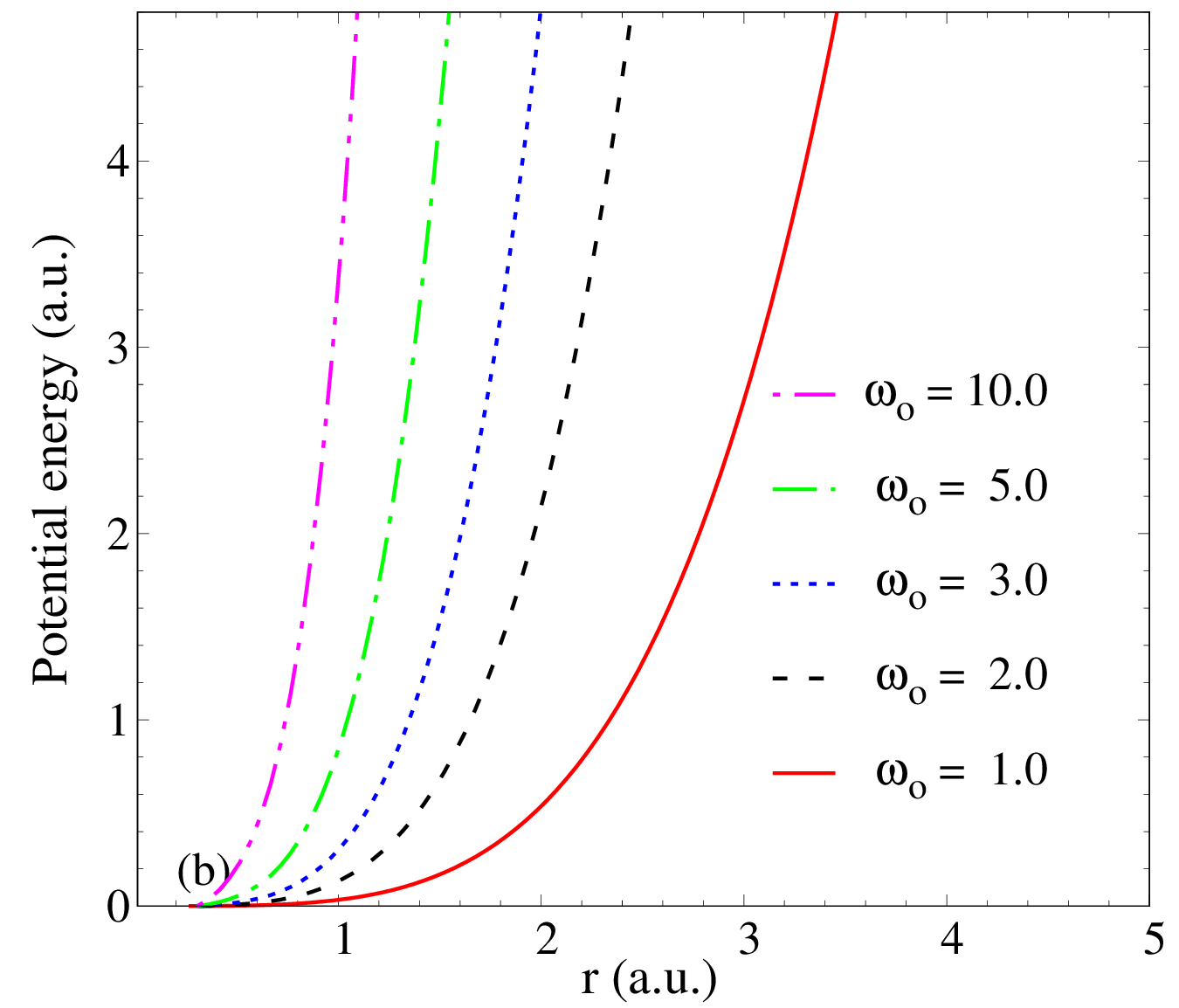}
} 
\quad 
{
\includegraphics[width=.4\textwidth]{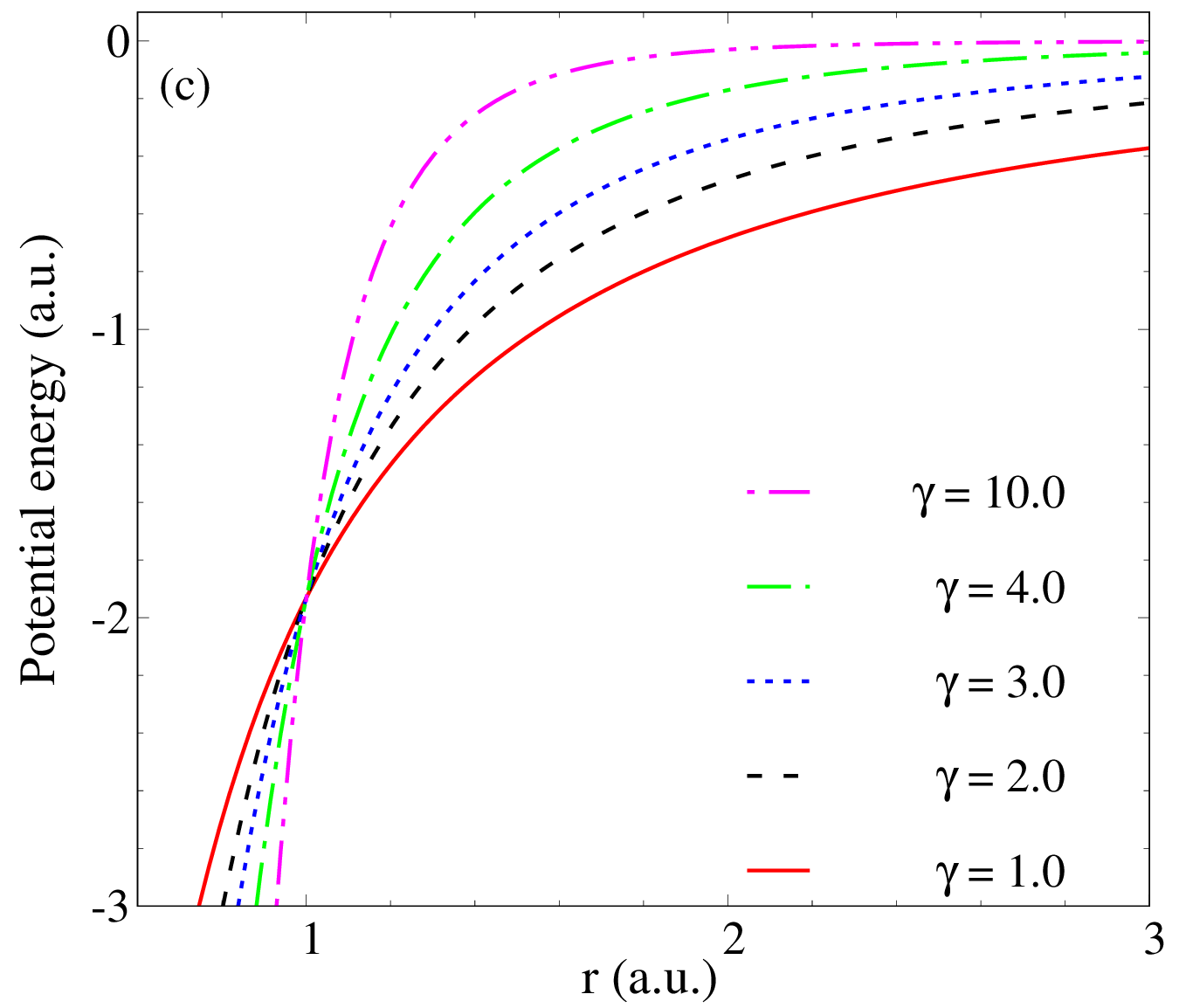}
}
\quad 
{
\includegraphics[width=.4\textwidth]{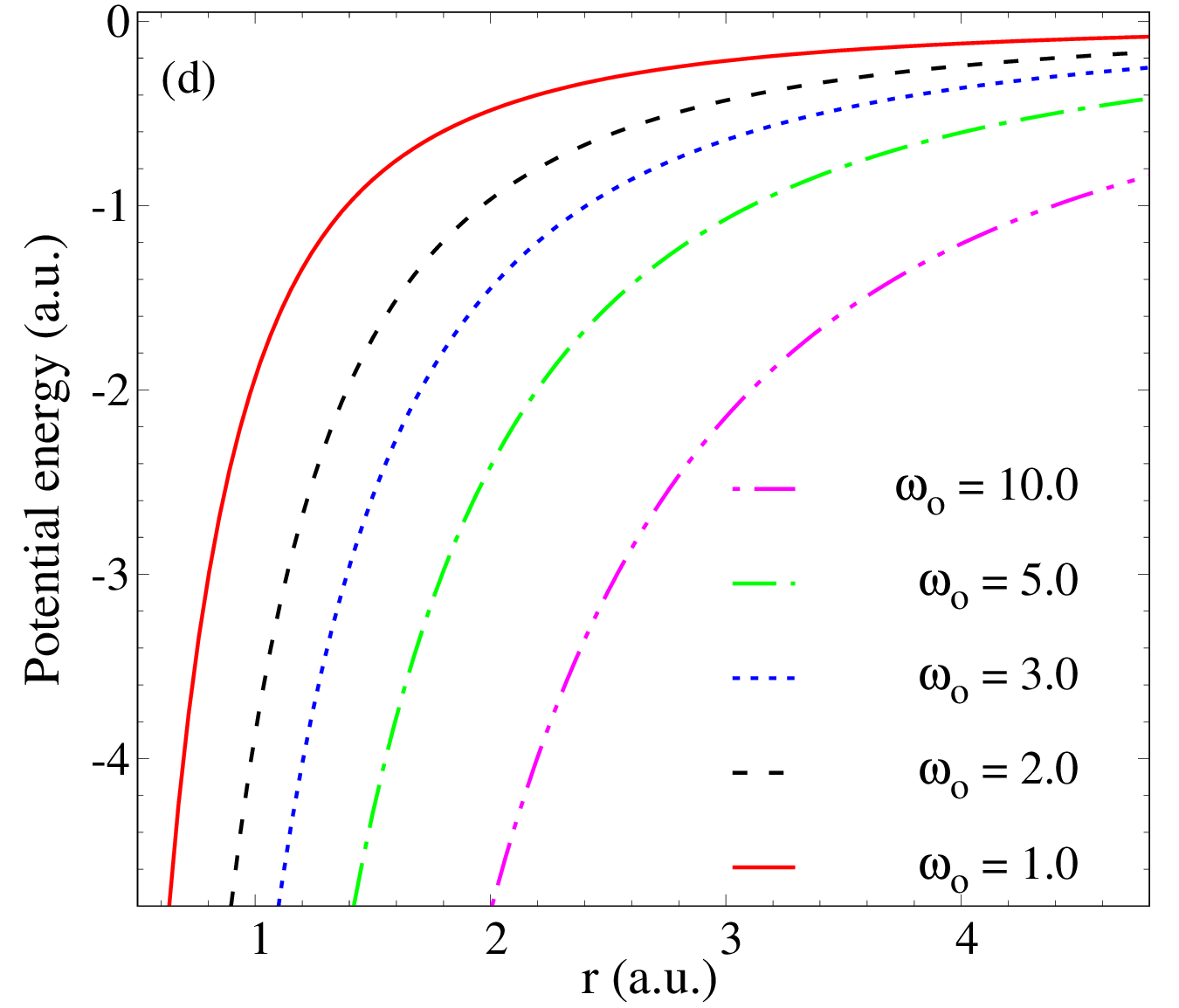}
} 
\caption{The a), and b)  quantum dot, and c), and d)  quantum antidot confinement potential as a function of distance for different values of the $\gamma$ ($\omega_{0}=1.0$), and $\omega_{0}$ ($\gamma=2.0$).}
\label{fig:potential_case1_case2}
\end{figure}

Before we delve deeper, it's essential to recognize how the confinement of electrons in a two-dimensional quantum ring can be characterized by a radial potential described as
\begin{equation}
\nu(r) = a_1 r^2 + \frac{a_2}{r^2} - \nu_0, \label{radial_potential}
\end{equation}
where $\nu_0 = 2\sqrt{a_1 a_2}$. At $r_0 = \left(a_1/a_2\right)^{1/4}$, this potential reaches its minimum, $\nu(r_0) = 0$, defining $r_0$ as the ring's average radius. Close to $r_0$, the potential assumes a nearly parabolic shape, $\nu(r) \approx \mu \omega_0^2 (r - r_0)^2/2$, where $\omega_0 = \sqrt{8a_2/\mu}$ represents the characteristic frequency, and $\mu$ denotes the effective mass of an electron in the semiconductor material. For a specified Fermi energy $E_F$, the effective width of the ring, $\Delta r$, is consequently defined as $\Delta r = \sqrt{8E_F/\mu \omega_0^2}$. By appropriately selecting the parameters $a_1$ and $a_2$, one can independently tailor the ring's radius and width to suit specific experimental or theoretical requirements. The potential (\ref{radial_potential}) describes, within certain limits, a quantum dot, a quantum antidot, a two-dimensional wire, or even a 1D ring \cite{tanInkson}. Although our case is 3D, we will evoke these definitions for quantum dots and anti-quantum dots. In the following, we will analyze variable mass models that induce variants of the quantum dot ($a_2\equiv0$) and the quantum antidot ($a_1\equiv0$). Figures \ref{fig:potential_case1_case2} demonstrate this fact. For a quantum dot, with $a_{1}\equiv \lambda^{4}/2$, we have an isotropic oscillator.
\begin{figure}[h!]
\centering
\includegraphics[scale=0.3]{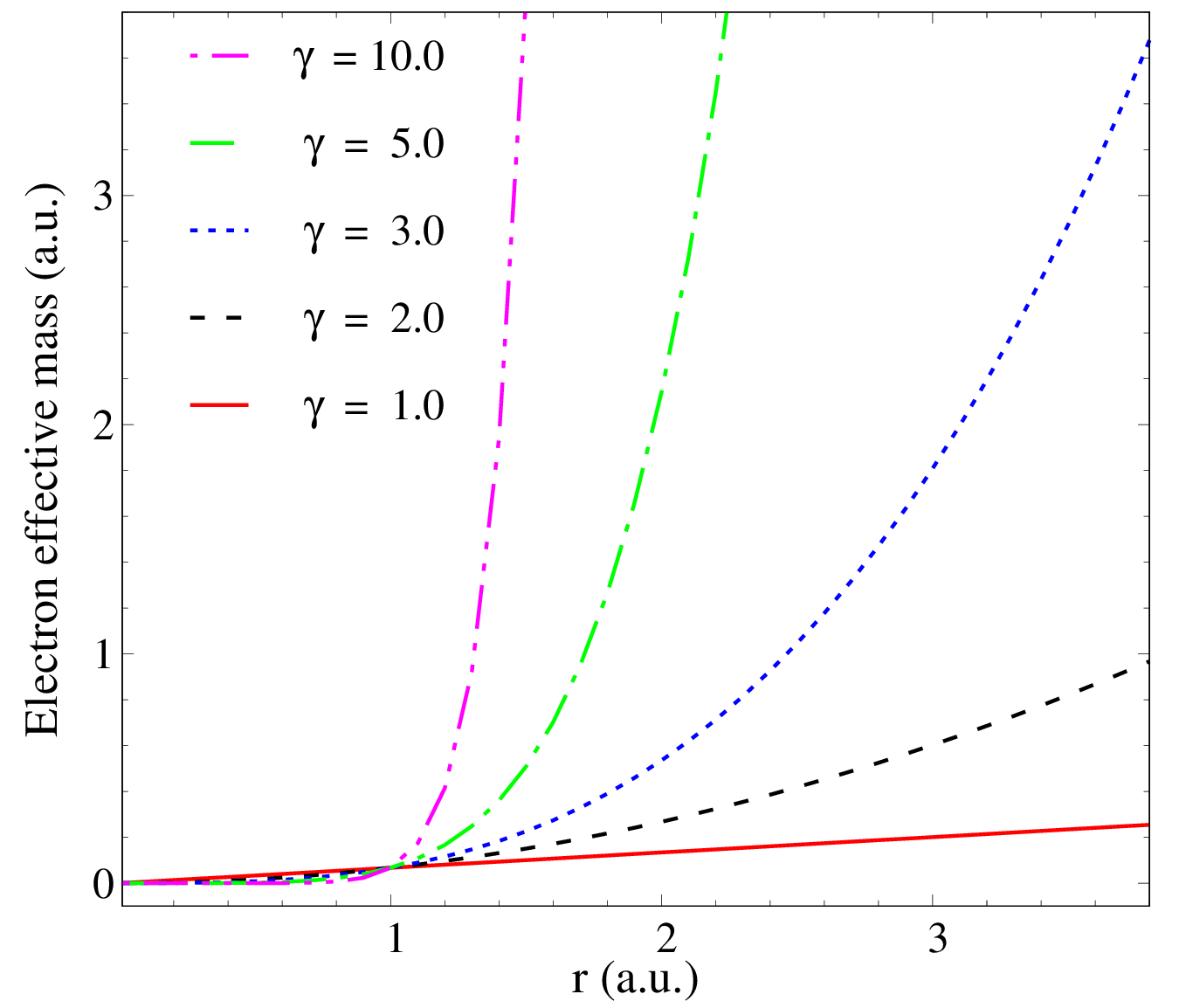}
\caption{The radial dependent mass for different values of the parameter $\gamma \neq 0$, and $\alpha=0.067m_{e}$, described in Eq (\ref{eq:mass}).}
\label{fig:spherical_mass_g}
\end{figure}

Our investigations consist of considering the radially variable mass \cite{Alhaidari} as 
\begin{equation}
m(r)=\alpha r^{\gamma}
\label{eq:mass}
\end{equation}
as showed in Figure \ref{fig:spherical_mass_g}, and taking the PCT function $q(r)=r^{\nu}$, where $\alpha$, $\gamma$, and $\nu$ are nonzero real parameters to obtain $V(r)$ and $E$ in Eq. (\ref{eq:final_equation}). By substitution in Eq. (\ref{eq:final_equation}), we obtain two solutions, which we will refer to as case 1 (a variant of the quantum dot) and case 2 (a variant of the quantum antidot), as described below.

\subsection{Case 1: quantum dots}

This case is obtained by imposing the following parameter constraints:
$\nu=1+\frac{\gamma}{2}$ and $\gamma\neq-2$. In this way, the potential $V(r)$ in Eq. (\ref{radial_potential}) reads \cite{Alhaidari}
$$V(r)=\frac{\alpha}{2}\omega_{0}^{2}r^{\gamma+2}.$$
For this particular case, the eigenvalues and the wave functions are found by Eq. (\ref{eq:final_equation}) and are given, respectively,
\begin{equation}
E_{n}=\left(\frac{\gamma+2}{2}\right)\left(2n+\Lambda+1\right)\omega_{0},
\end{equation}
and
\begin{equation}
   \phi_{n}(r)=A_{n}(\xi r)^{\left(1+\frac{\gamma}{2}\right)\Lambda+\frac{\left(1+\gamma\right)}{2}}\exp\left[-\frac{\left(\xi r\right)^{\gamma+2}}{2}\right]L_{n}^{\Lambda}\left[\left(\xi r\right)^{\gamma+2}\right]\;, 
\end{equation}
where $\Lambda(l)=|\gamma+2|^{-1}\sqrt{4l(l+1)+(\gamma-1)^{2}}$, $\omega_{0}$ is a real potential coupling parameter.

\subsection{Case 2: quantum antidots}

In this case, the imposed parameter constraints are
$\nu=1+\frac{\gamma}{4}$ and $\gamma\neq-2$, yielding \cite{Alhaidari}
\begin{equation}
V(r)=-\frac{\omega_{0}}{2\sqrt{\alpha}}r^{-1-\frac{\gamma}{2}}.
\end{equation}
For this case, the eigenvalues are given by
\begin{equation}
E_{n}=-\frac{\omega_{0}^{2}}{2\left(\gamma+2\right)^{2}}\frac{1}{\left(n+\Lambda+\frac{1}{2}\right)^{2}}\;,
\end{equation}
while the wave functions are described by
\begin{equation}
\phi_{n}(r)=A_{n}(\xi_{n}r)^{\left(1+\frac{\gamma}{2}\right)\Lambda+\frac{\left(1+\gamma\right)}{2}} \exp\left[- \frac{\left(\xi_{n}r\right)^{\gamma+2}}{2}\right] L_{n}^{2\Lambda}\left[\left(\xi_{n}r\right)^{1+\frac{\gamma}{2}}\right]\;,
\end{equation}
where $$\Lambda(l)=|\gamma+2|^{-1}\sqrt{4l(l+1)+(\gamma-1)^{2}},$$ and
$$\xi_{n}=\frac{4\sqrt{\alpha}\omega_{0}}{\left(\gamma+2\right)^{2}\left(n+\Lambda+\frac{1}{2}\right)^{\frac{1}{1+\frac{\gamma}{2}}}}.$$
Figure \ref{fig:potential_case1_case2} shows two conditions for quantum dot and quantum antidot potential.

\section{\label{sec:optics}Optical properties}

In this section, we shall investigate optical properties in quantum dot and antidot systems, specifically the OAC and RIC. These properties are critical for understanding the interaction between incident electromagnetic waves and the confined electronic states within quantum systems. The study encompasses the linear and nonlinear contributions to OAC and RIC, emphasizing intersubband transitions driven by the dipole selection rules. These optical phenomena play a crucial role in designing and optimizing optoelectronic devices, as they influence the efficiency of energy conversion processes in quantum systems. By examining the variations in these optical properties as functions of photon energy and system parameters, we aim to comprehensively analyze how quantum confinement and effective mass distribution shape the behavior of quantum dots and antidots. Following this, we calculate the OAC and RIC for both quantum dot and antidot structures, considering the intersubband transitions between two subbands based on the selection rules. We also investigate the influence of different system parameters on the resulting optical responses.

Following the references \cite{rosencher,ahn}, the OAC and RIC are calculated for the quantum dot (antidot) structure considering the intersubband transitions between two subbands. For our purpose, the electric dipole selection rules produce $\Delta l = \pm 1$ thus, $E_{21}= E_{2}-E_{1}$ is the energy interval of the two-level systems, and $\psi_{2}=\psi_{n=0,l=0}$, and $\psi_{2}=\psi_{n=1,l=1}$ are the correspondent wavefunctions.

Assuming the polarization of the incident radiation along the in-plane $x$-axis, the intrasubband dipole matrix elements are
\begin{equation}
M_{21}=\langle \psi_{2}|{e r \cos{\theta}}|\psi_{1}\rangle,\label{M12}
\end{equation}
with the diagonal values of the matrix elements $M_{11}=M_{22}=0$, due to the azimuthal symmetry of the system. Considering (\ref{M12}), the analytical forms of the linear and the nonlinear OAC are given, respectively, by
\begin{equation}
\alpha ^{(1)}\left( \omega \right) =\hslash \omega \sqrt{\frac{\mu_0 }{
\epsilon _{r}}}\frac{\sigma _{\nu }\Gamma
_{21}\left\vert M_{21}\right\vert ^{2}}{\left(E_{21}-\hslash \omega \right) ^{2}+\left( \hslash \Gamma
_{21}\right) ^{2}},  \label{a1}
\end{equation}
and
\begin{equation}
\alpha ^{(3)}\left( \omega ,I\right) =-\hslash \omega \sqrt{\frac{\mu_0}{
\epsilon_{r}}}\frac{4 I}{2\epsilon _{0}n_{r}c}\frac{\sigma _{\nu }\Gamma
_{21}\left\vert M_{21}\right\vert ^{4}}{\left[\left(E_{21}-\hslash
\omega \right) ^{2}+\left( \hslash \Gamma _{21}\right) ^{2}\right]^2}.
\end{equation}

The total OAC can be written as
\begin{equation}
\alpha \left( \omega ,I\right) =\alpha ^{(1)}\left( \omega \right) +\alpha
^{(3)}\left( \omega ,I\right).  \label{at}
\end{equation}

We also obtained the linear, nonlinear, and total RIC, respectively as
\begin{equation}
\frac{\Delta n^{(1)}\left( \omega\right) }{n_r} =\frac{\sigma _{\nu }\left\vert M_{21}\right\vert ^{2}}{2 n_r^{2}\epsilon_0}\frac{E_{21}-\hslash\omega}{\left(E_{21}-\hslash \omega \right) ^{2}+\left( \hslash \Gamma
_{21}\right) ^{2}},  \label{n1}
\end{equation}
\begin{equation}
\frac{\Delta n^{(3)}\left( \omega ,I\right)}{n_r} =-\frac{ \mu c I \sigma _{\nu }\left\vert M_{21}\right\vert ^{4}}{\epsilon _{0}n_{r}^3}\frac{E_{21}-\hslash\omega}{\left[\left(E_{21}-\hslash
\omega \right) ^{2}+\left( \hslash \Gamma _{21}\right) ^{2}\right]^2}, \label{n3}
\end{equation}
and
\begin{equation}
\frac{\Delta n \left( \omega ,I\right)}{n_r} =\frac{\Delta n^{(1)}\left( \omega\right) }{n_r}+\frac{\Delta n^{(3)}\left( \omega ,I\right)}{n_r}.  \label{nt}
\end{equation}

\begin{figure}[tbh]
\centering
\includegraphics[scale=0.3]{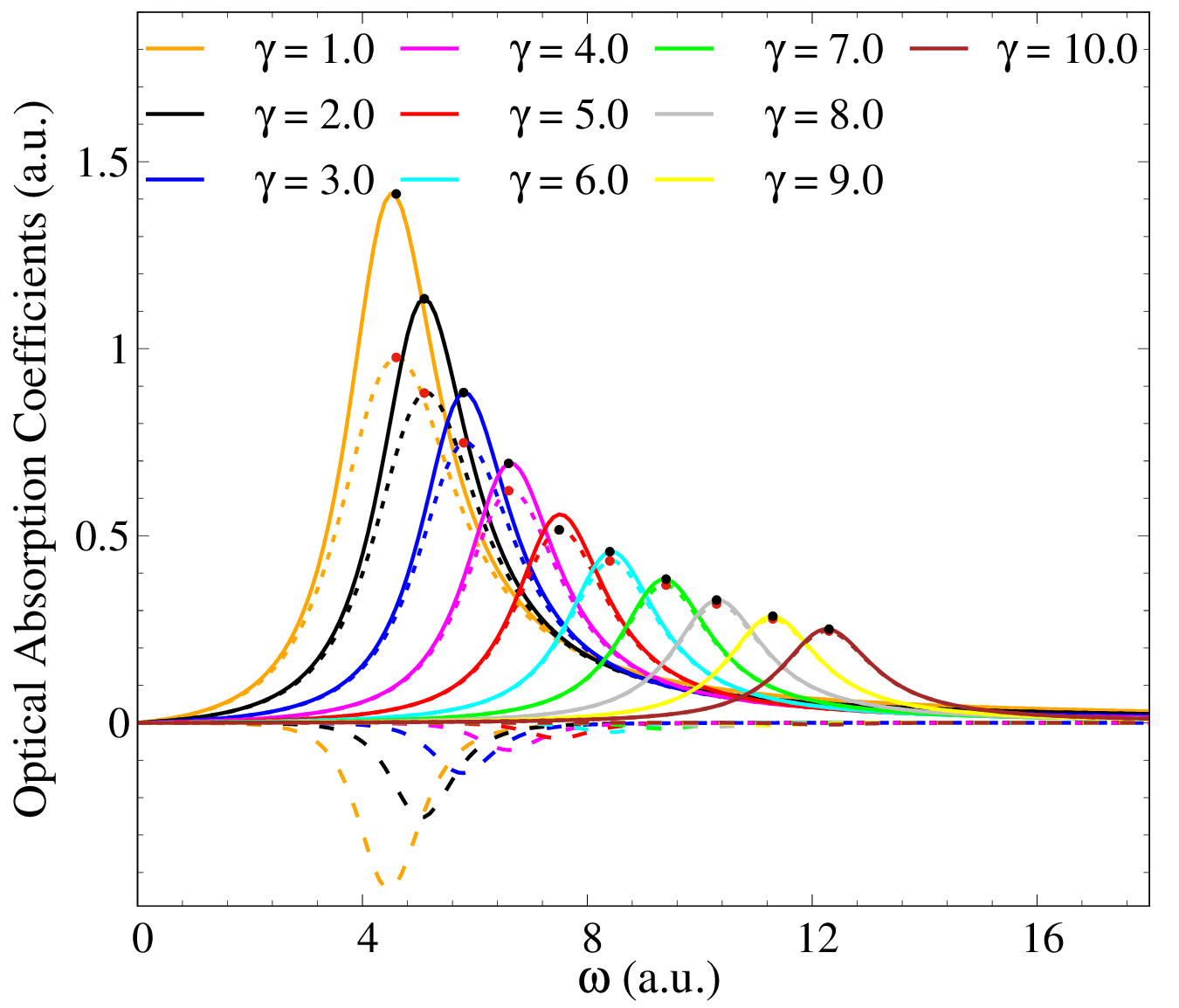}
\caption{The linear (solid line), third-order non-linear (dashed line), and total (dotted line) OAC as a function of the photon energy for different values of the parameter $\gamma$. The points suggested an exponential decay of the linear OAC.}
\label{fig:optical_case1_g}
\end{figure}

\begin{figure*}[h!]
\centering
{
\includegraphics[width=.4\textwidth]{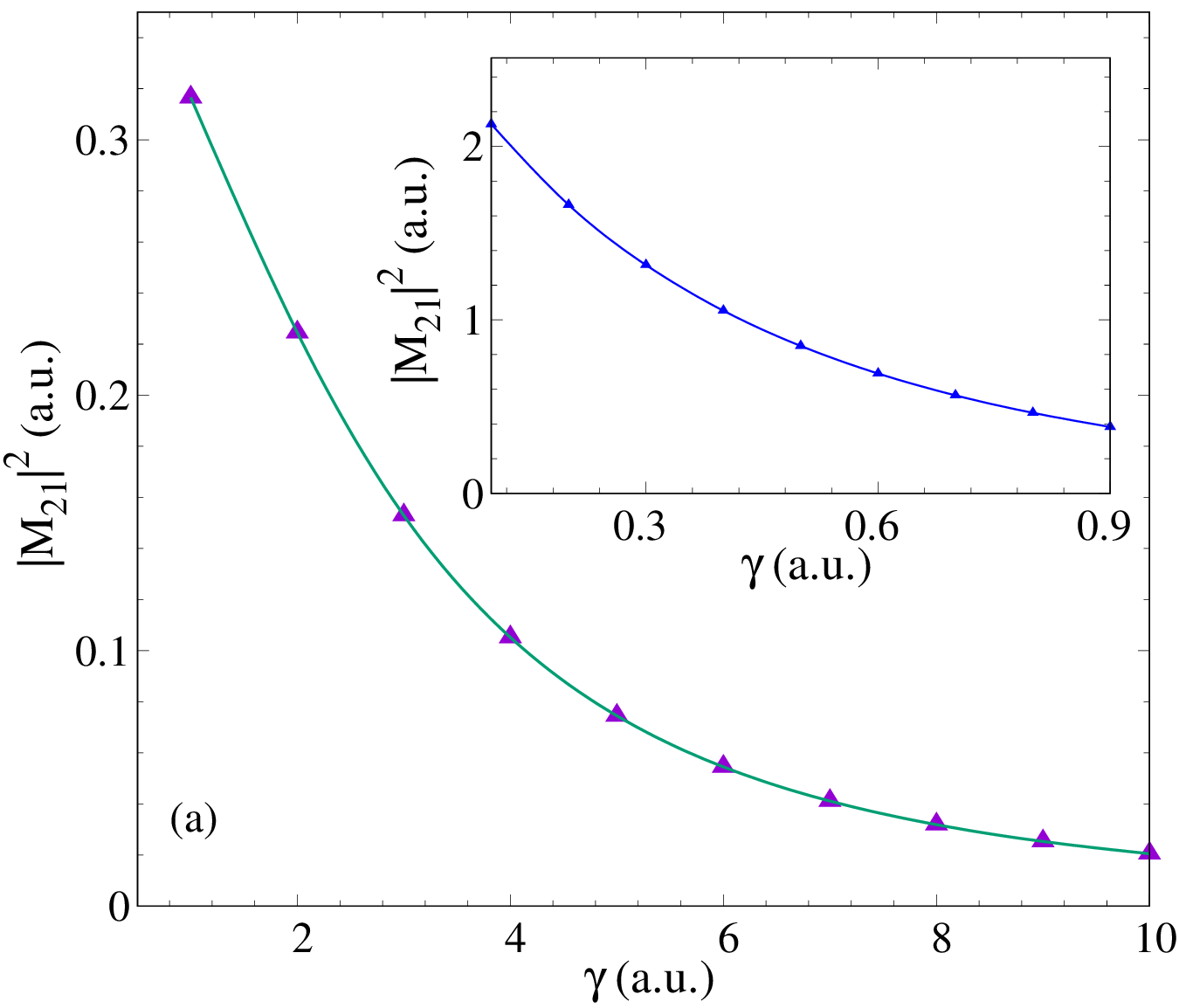}
} 
\quad 
{
\includegraphics[width=.4\textwidth]{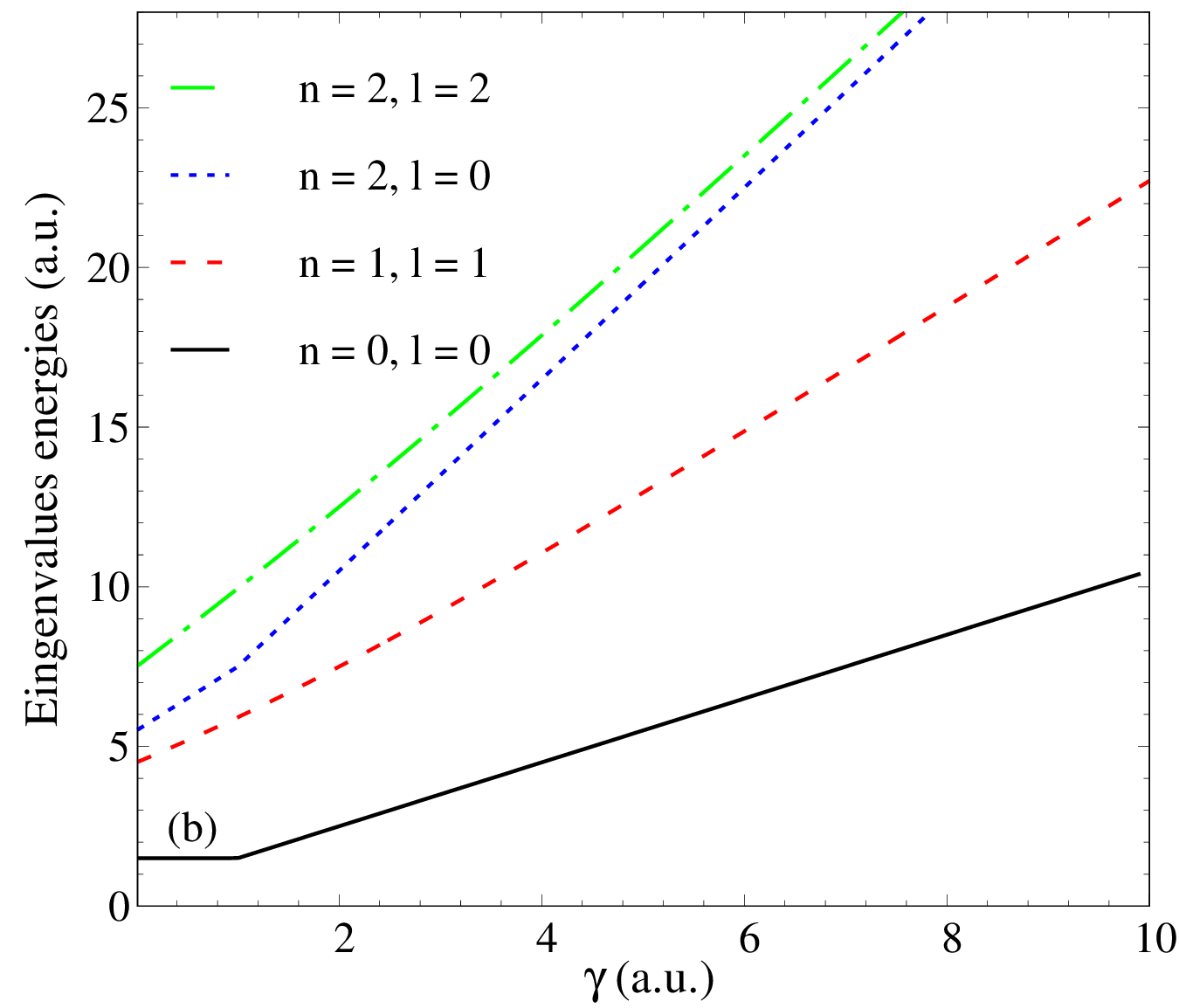}
} 
\caption{(a) Intrasubband dipole matrix elements, and (b) energy spectrum as a function of the parameter $\gamma$ and $\omega_{0}=1.0$. The insert stands for the small values of $\gamma$.}
\label{fig:combo_case1_g}
\end{figure*}

\section{\label{sec:results}Results and discussions}

In this section, we study the optical properties addressed in Sec. \ref{sec:optics} for the model presented in Sec. \ref{sec:PCT}. We discuss the situations that lead to the analog of a quantum dot and an anti-quantum dot originating directly from the PDM. In our graphical analyses, we use the parameters written in atomic units, $c=\hbar=m_{e}=e=1$.

\subsection{\label{sec:case1_results}Case 1: quantum dots}

In this subsection, we examine the optical properties of quantum dots with a position-dependent mass distribution. The focus is on understanding the role of the confinement parameter $\omega_0$ and the mass scaling parameter $\gamma$ in shaping the linear and nonlinear OAC and the RIC. Through the variation of these parameters, we investigate the shifting of energy levels, the intrasubband dipole matrix elements, and the resulting optical phenomena that arise in quantum dots. We illustrate the optical characteristics of quantum dots, highlighting the dependency of OAC and RIC on different values of $\gamma$ and $\omega_0$. Figure \ref{fig:optical_case1_g} shows the optical absorption coefficient, $\alpha ^{(1)}(\omega)$, $\alpha ^{(3)}(\omega, I)$, and $\alpha(\omega, I)$ as a function of incident photon energy $\hbar\omega$, for different values of parameter $\gamma$, with coupling parameter of the potential, $\omega_{0}=1.0$, and $I=0.5$. As observed, the resonance peak of absorption is strongly related to the parameter $\gamma$. As the value of $\gamma$ increases, the peaks decrease monotonically while simultaneously shifting towards higher energies.

\begin{figure}[!t!]
\centering
\includegraphics[scale=0.45]{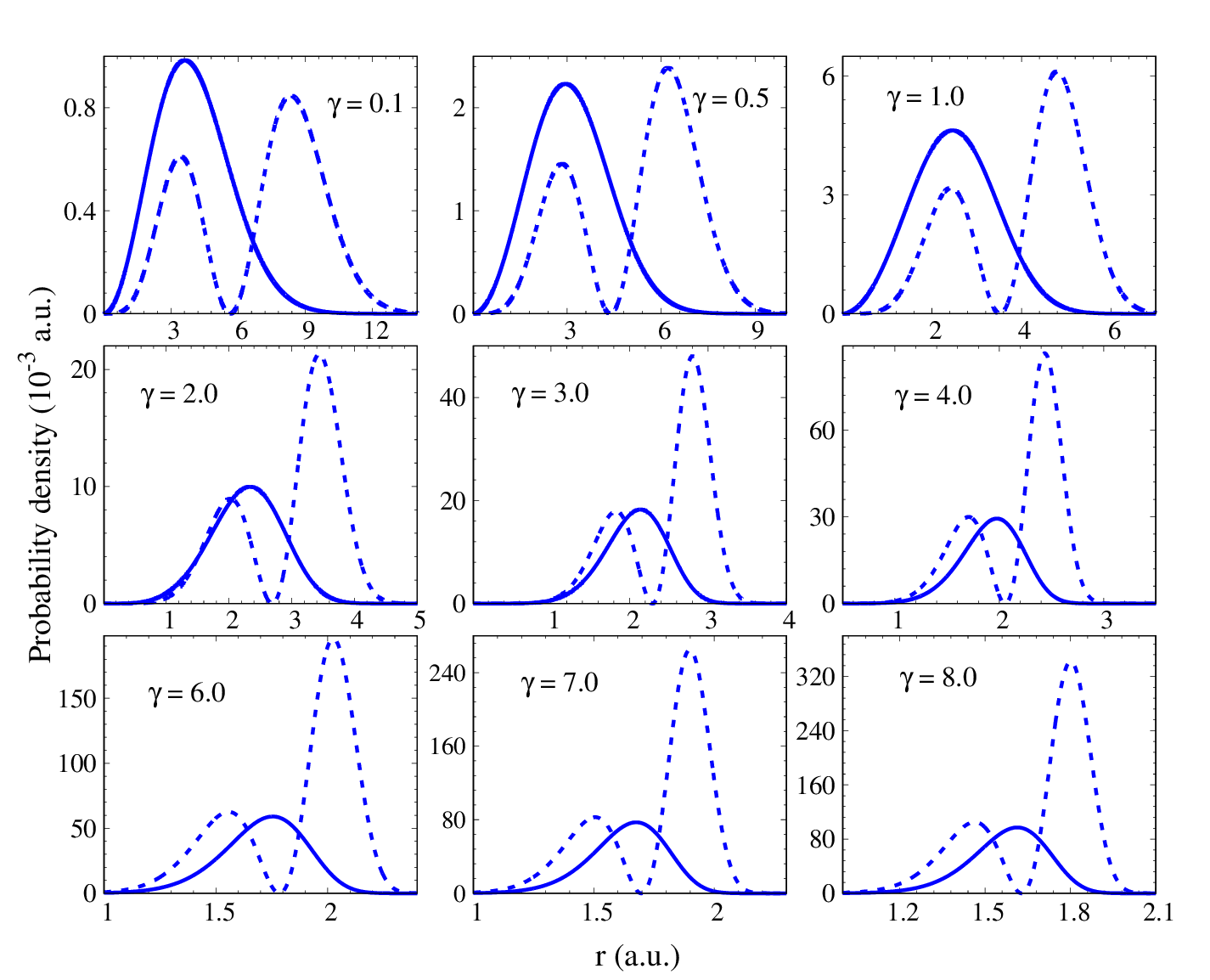}
\caption{The probability density of the quantum dots states participating in the transition: solid line for the ground state ($n=0,l=0$), and dashed-line for the first excited state ($n=1,l=1$), as a function of the parameter $\gamma$, and $\omega_{0}=1.0$.}
\label{fig:wavefunction_g_case1}
\end{figure}

\begin{figure}[!t!]
\centering
\includegraphics[scale=0.3]{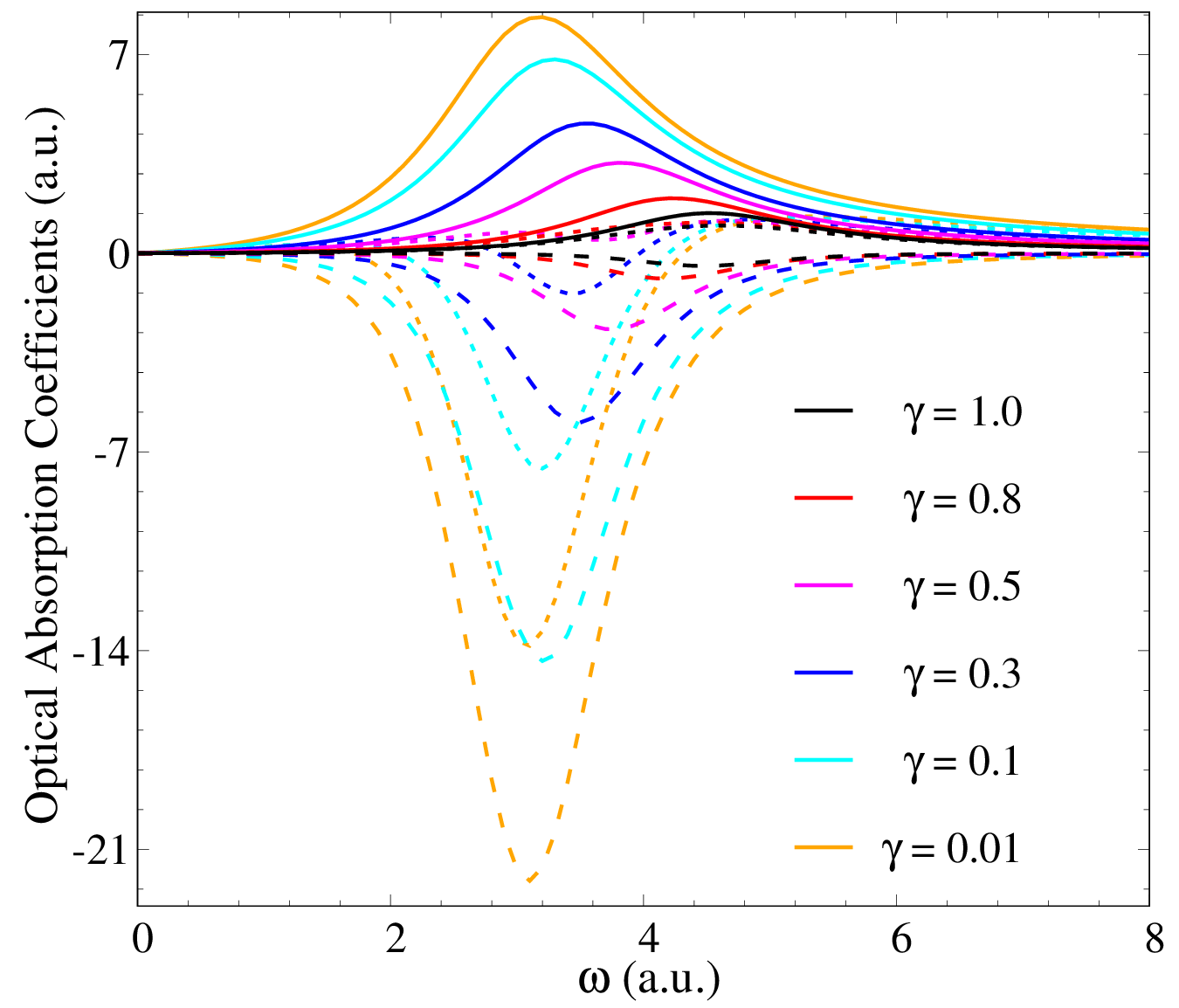}
\caption{ The linear (solid line),  third-order non-linear (dashed line), and total (dotted line) absorption coefficient of the quantum dot as a function of the photon energy for small values of the parameter $\gamma$, and $\omega_{0}=1.0$.}
\label{fig:variable_small_g}
\end{figure}

These features could be explained by analyzing how the difference between the first energy levels, $E_{21}$, and the intrasubband dipole matrix elements, $|M_{21}|$, vary with that parameter, as shown in Figure \ref{fig:combo_case1_g}. As noted, $|M_{21}|$ decreases with the increase of $\gamma$, towards values close to zero, at the same time $E_{21}$ increases with increasing $\gamma$.   This is expected since the potential becomes more and more confining to larger $\gamma$, as shown in Fig. \ref{fig:potential_case1_case2}(a), thereby explaining the redshift of the peak of the optical absorption coefficient.

\begin{figure}[!t]
\centering
\includegraphics[scale=0.3]{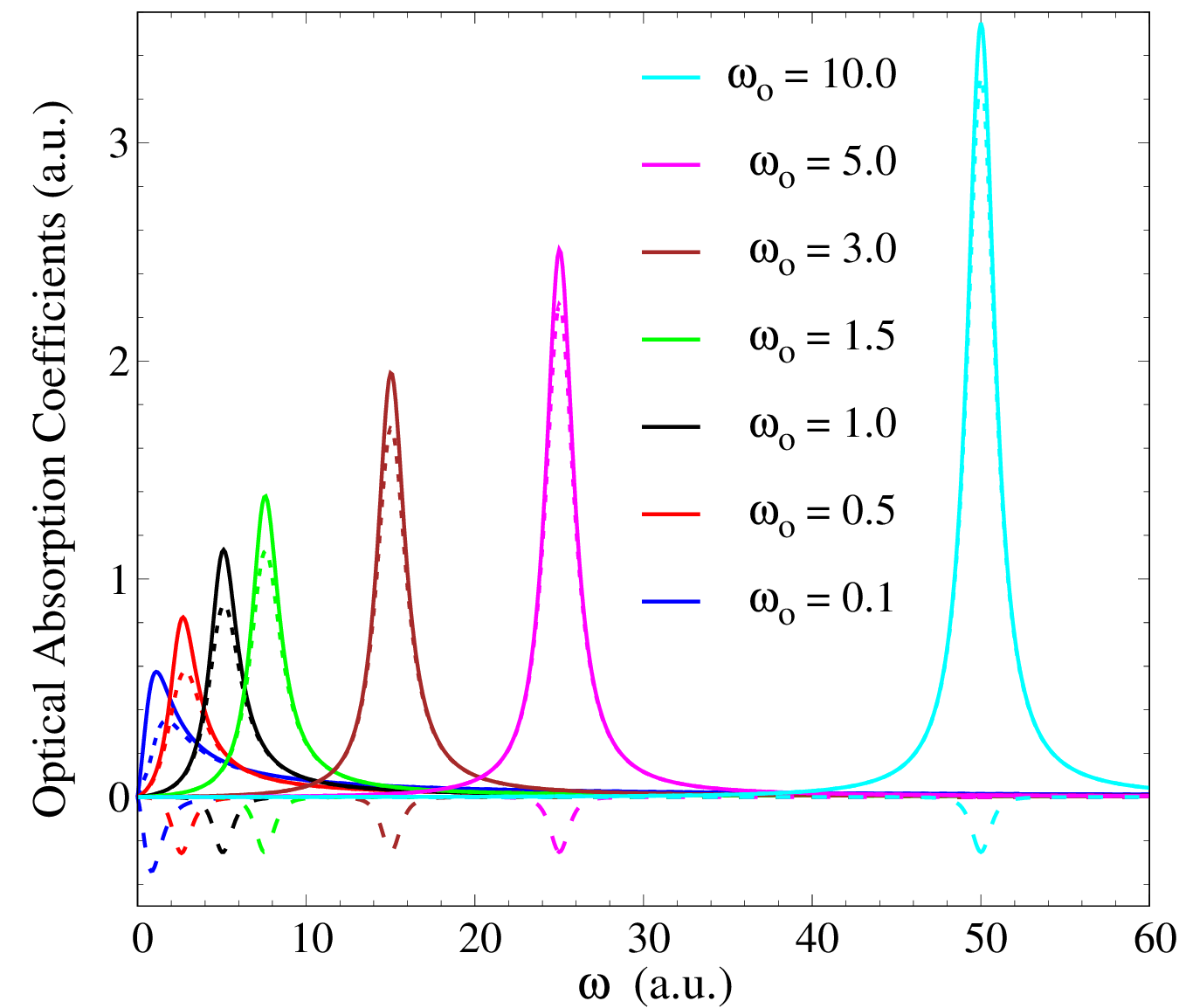}
\caption{The linear (solid line), third-order non-linear (dashed line), and total absorption coefficient (dotted line) as a function of the photon energy for different values of the parameter $\omega_{0}$ of the quantum dot potential.}
\label{fig:optical_case1_c}
\end{figure}

\begin{figure*}[!t]
\centering
{
\includegraphics[width=.4\textwidth]{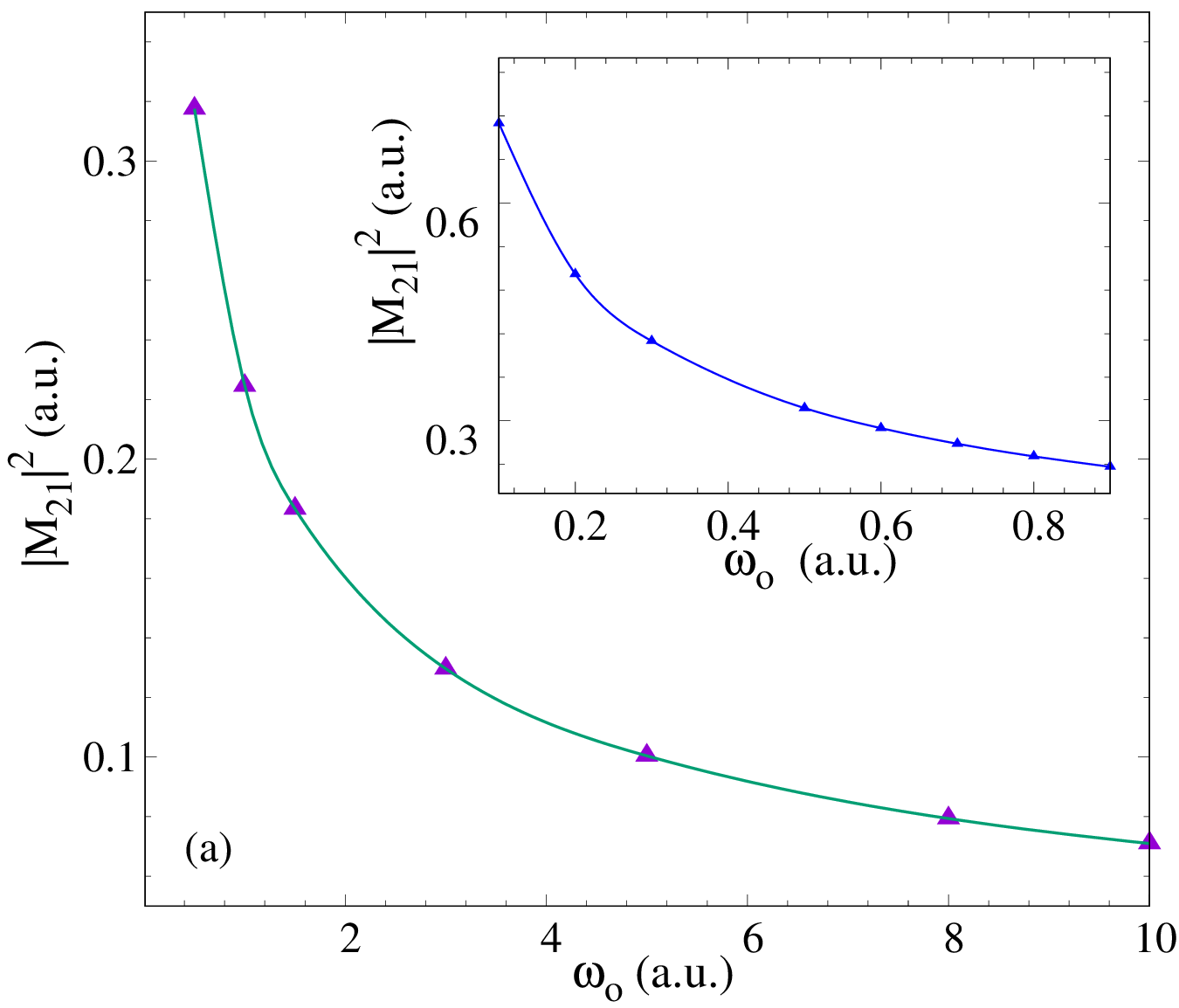}
} 
\quad 
{
\includegraphics[width=.4\textwidth]{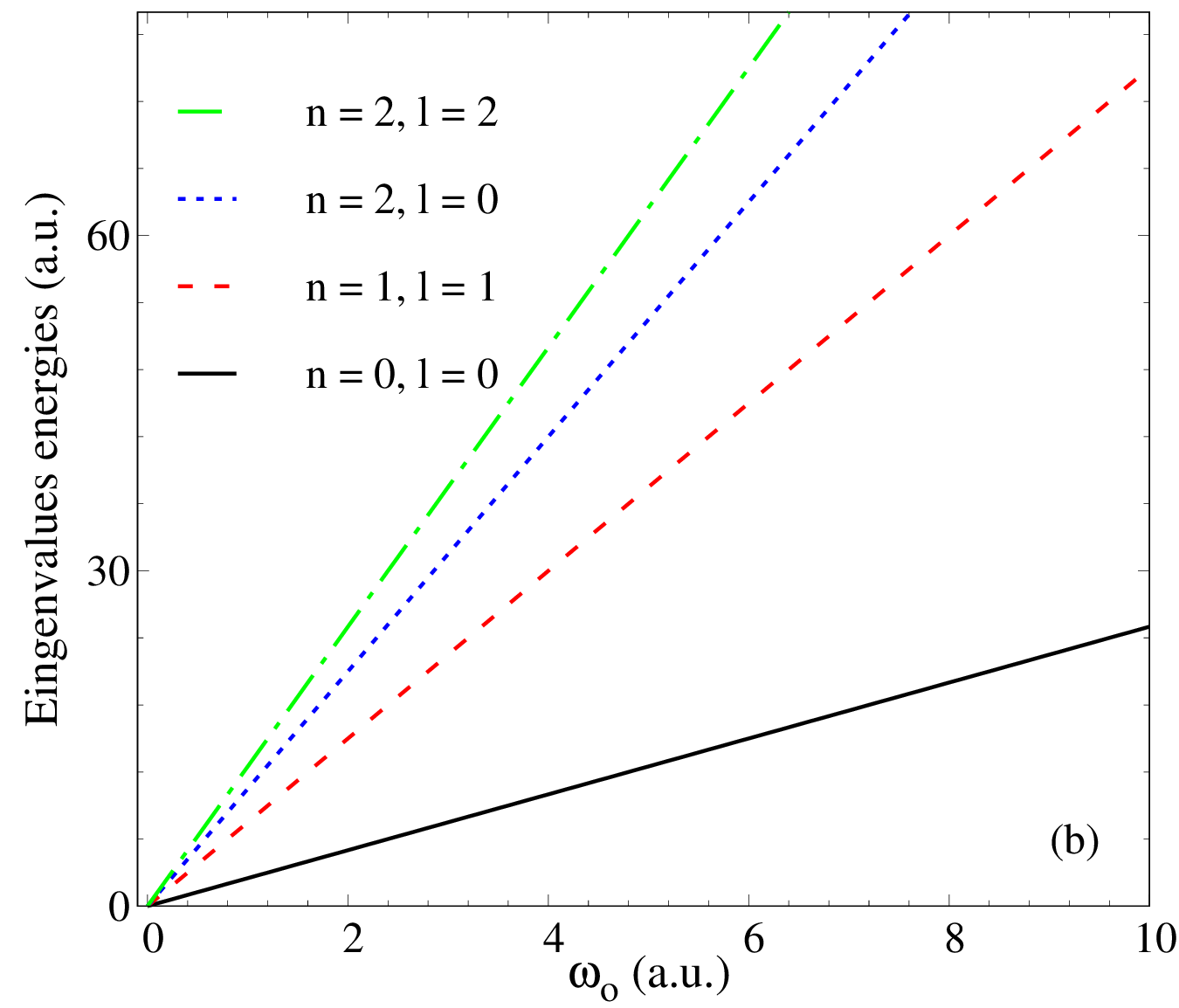}
} 
\caption{(a) Intrasubband dipole matrix elements, and (b) energy spectrum as a function of the parameter $\omega_{0}$ and $\gamma=2.0$ of the quantum dot potential. The insert represents the small $\omega_{0}$ values.}
\label{fig:combo_case1_c}
\end{figure*}

\begin{figure}[!t]
\centering
\includegraphics[scale=0.45]{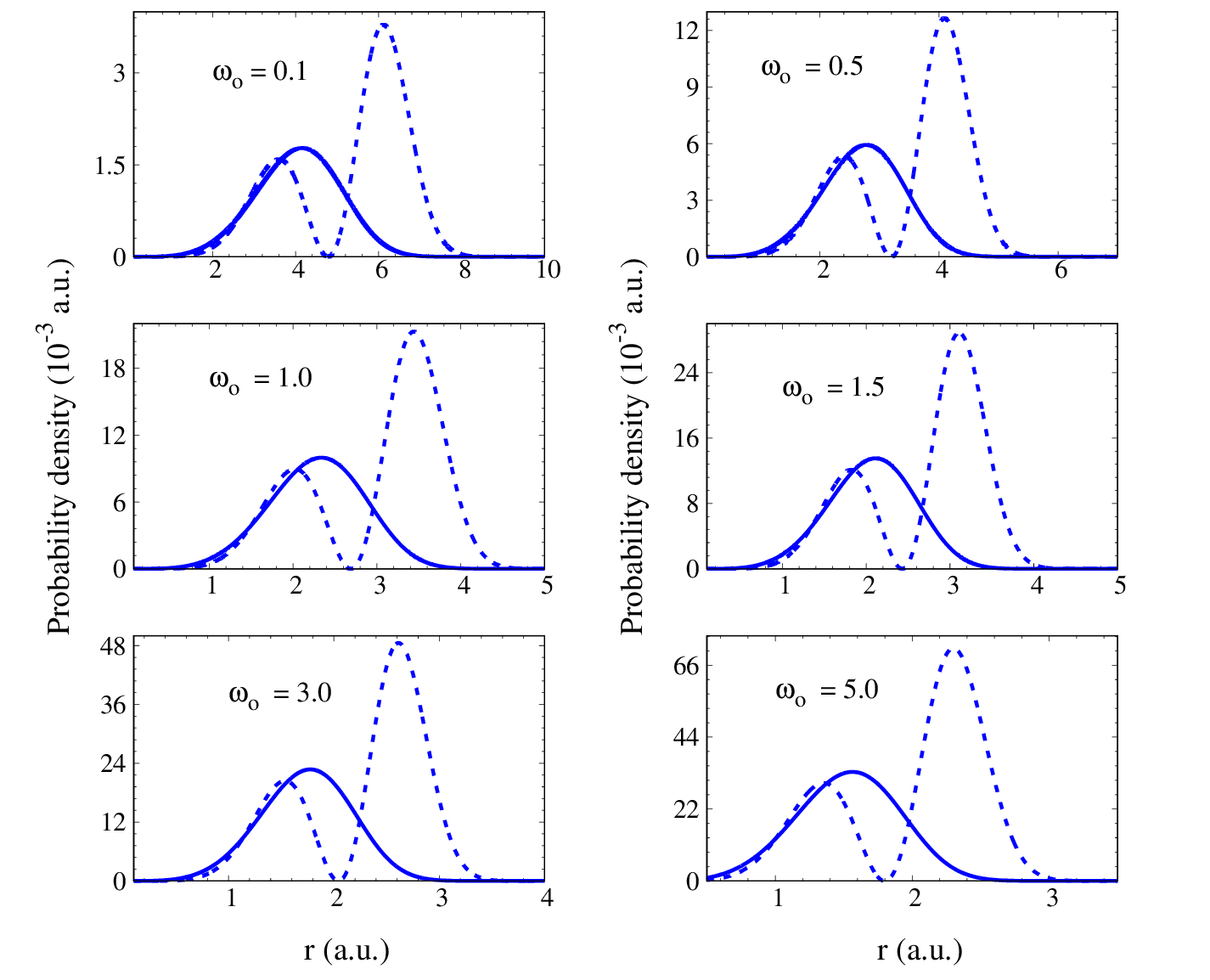}
\caption{The probability density of the quantum dot states participating in the transition: solid line for the ground state ($n=0,l=0$), and dashed-line for the first excited state ($n=1,l=1$), as a function of the parameter $\omega_{0}$, and $\gamma=2.0$.}
\label{fig:wavefunction_c_case1}
\end{figure}

Conversely, when $\gamma$ approaches zero, the third-order non-linear contribution of the absorption coefficient becomes significant, leading to a negative total absorption coefficient, as seen in Fig. \ref{fig:variable_small_g}. The diminishing $\gamma$ produces a considerable increase of $|M_{21}|$ by an enhancement of the wavefunction overlapping, showing in Fig. \ref{fig:wavefunction_g_case1}, added to the $ E_{21}$ diminishing, as seen in Fig. \ref{fig:combo_case1_g}(b), generating the OAC shown in Fig. \ref{fig:variable_small_g}. 
In contrast, Fig. \ref{fig:optical_case1_c} shows the optical absorption coefficient for different values of confinement frequency $\omega_{0}$, with  $\gamma=2.0$, and $I=0.5$. As noted in Fig. \ref{fig:optical_case1_c},  the $\alpha^{(1)}(\omega)$, $\alpha ^{(3)}(\omega, I)$, and $\alpha(\omega, I)$ present a remarkable increase of the resonances peaks, and its position shows a considerable red shift when the frequency $\omega_{0}$ raises. In this case, the values of the $|M_{21}|$ decay smoothly, as presented in Fig. \ref{fig:combo_case1_c}(a) while the subbands energy interval, $E_{21}$, exhibits a rapid growth as reported in Fig. \ref{fig:combo_case1_c}(b), compared to earlier data (see Fig. \ref{fig:combo_case1_g}(b)), contributing to a sharp rise of the absorption optical coefficients. On the other hand, smaller values of the parameter $\omega_{0}$ results in a more considerable nonlinear contribution to the optical absorption coefficient, as shown in Fig. \ref{fig:optical_case1_c}, due principally to the sizeable overlapping region of the wavefunction, as seen in Fig. \ref{fig:wavefunction_c_case1}, resulting in larger values of the $|M_{21}|$, inset of the Fig. \ref{fig:combo_case1_c}(a).

\begin{figure}[!t]
\centering
\includegraphics[scale=0.3]{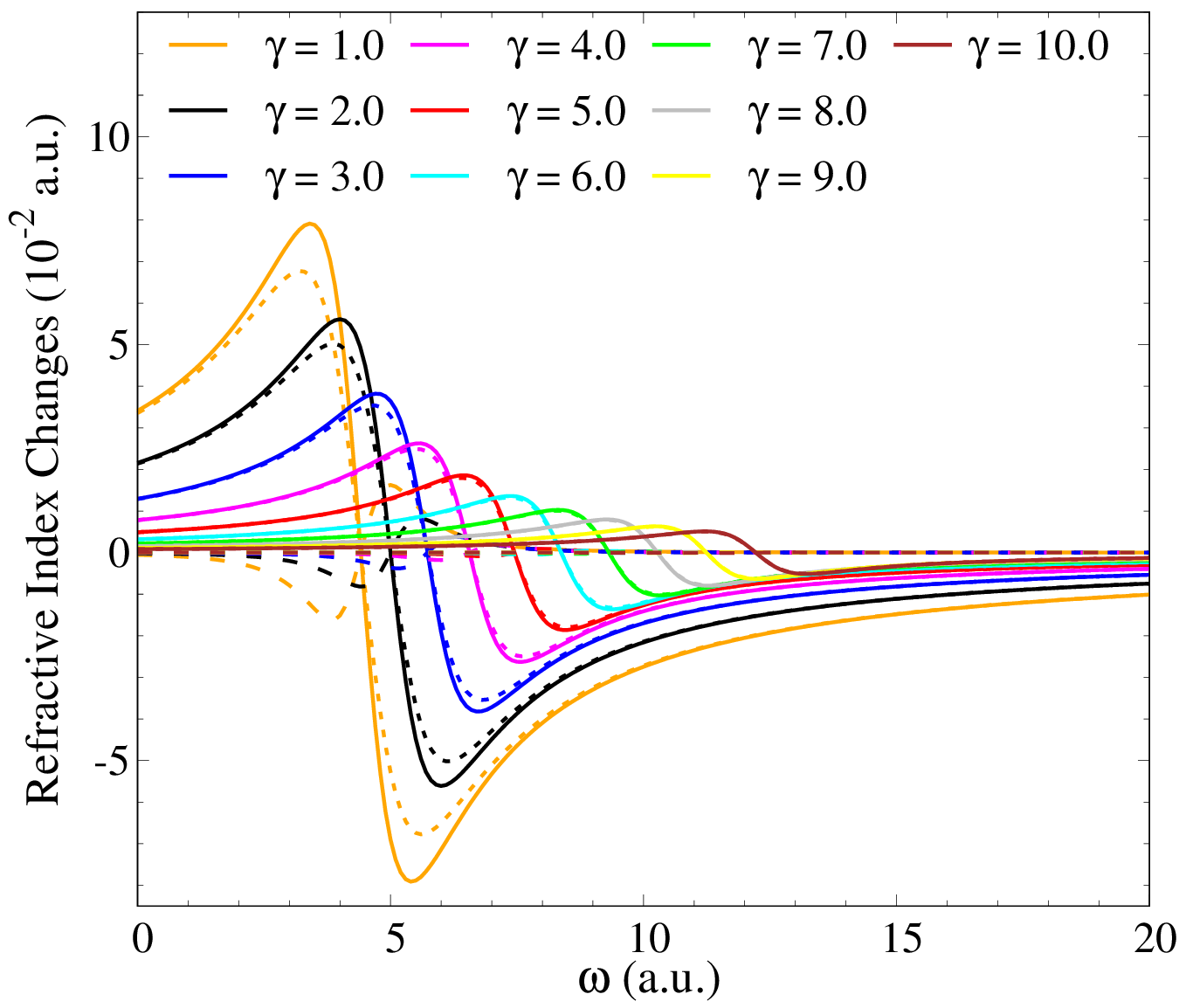}
\caption{The linear (solid line), third-order non-linear (dashed line), and total refractive index (dotted line) of the quantum dot as a function of the photon energy for different values of the parameter $\gamma$, $\omega_{0}=1.0$, and $I=0.5\, \text{(a.u.)}$.}
\label{fig:refractive_index_g}
\end{figure}

\begin{figure}[!t]
\centering
\includegraphics[scale=0.3]{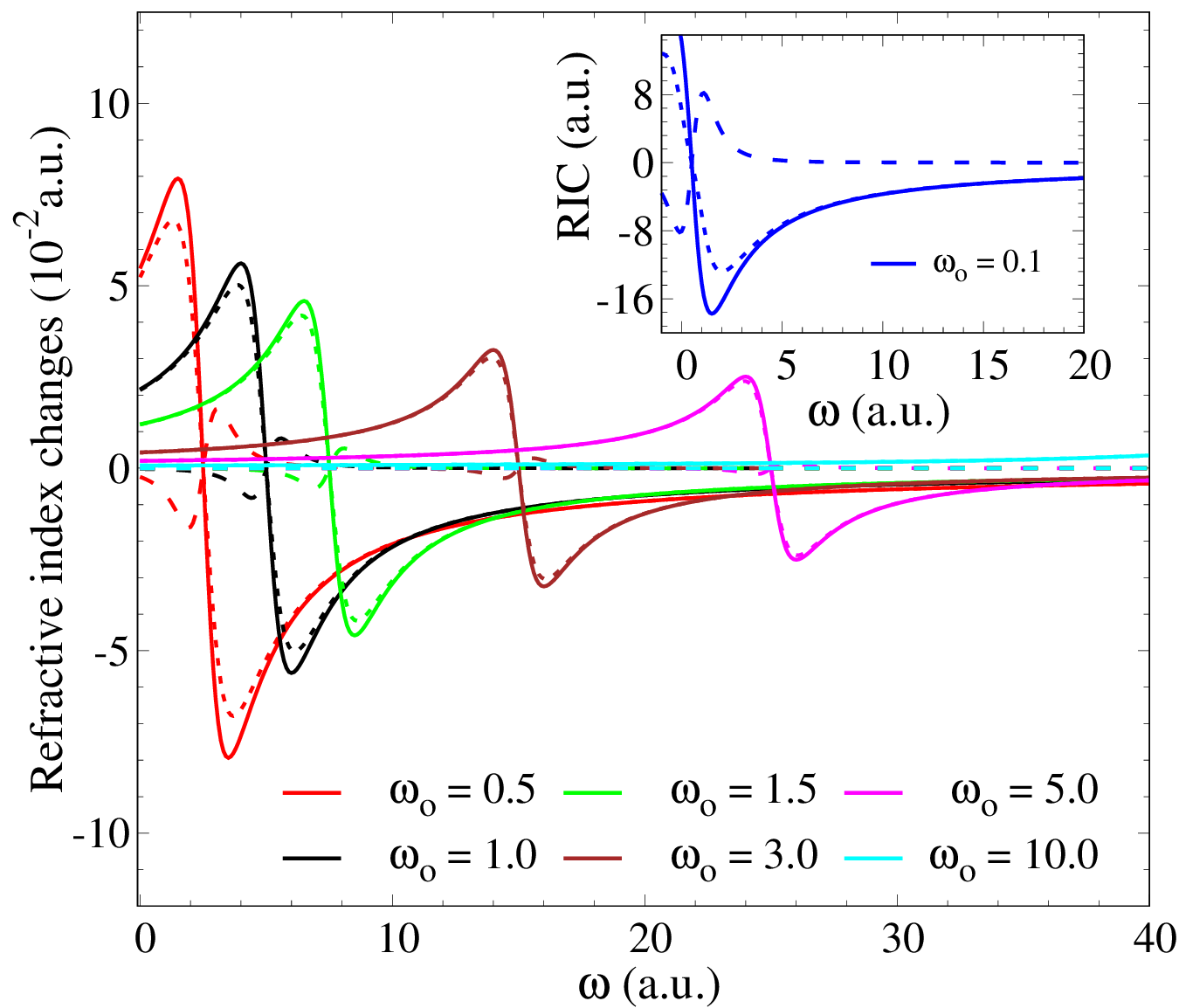}
\caption{The linear (solid line), third-order non-linear (dashed line), and total refractive index changes (dotted line) of the quantum dot as a function of the photon energy for different values of the parameter $\omega_{0}$, $\gamma=2.0$, and $I=0.5 \,\text{(a.u.)}$. The insert shows the refractive index changes for small values of $\omega_{0}$.}
\label{fig:refractive_index_c}
\end{figure}

Figure \ref{fig:refractive_index_g} describes the variations of the linear, $\Delta n^{(1)}\left( \omega\right)/n_r$,  nonlinear, $\Delta n^{(3)}\left(\omega, I\right)/n_r$, and total, $\Delta n \left(\omega, I\right)/n_r$ RIC as function of the incident energy of the photon for different values of $\gamma$. We observe that the RIC increases with the incident photon energy and shifts its peaks toward higher energies with the increase of $\gamma$. Besides, the resonance peak decreases with the increasing $\gamma$. In turn, the variation of the oscillator frequency, $\omega_{0}$, causes an abrupt shift towards higher energies, as shown in Fig. \ref{fig:refractive_index_c}, while the resonance peaks diminish with higher values of the $\omega_{0}$. On the other hand, to small values of the parameter $\omega_{0}$, the nonlinear RIC is the opposite in sign of the linear RIC, reducing the total RIC, as seen in the inset of the Fig.\ref{fig:refractive_index_c}. We observe a decrease in the nonlinear contribution to the refractive index change in both cases.

\begin{figure*}[!t]
\centering
{
\includegraphics[width=.4\textwidth]{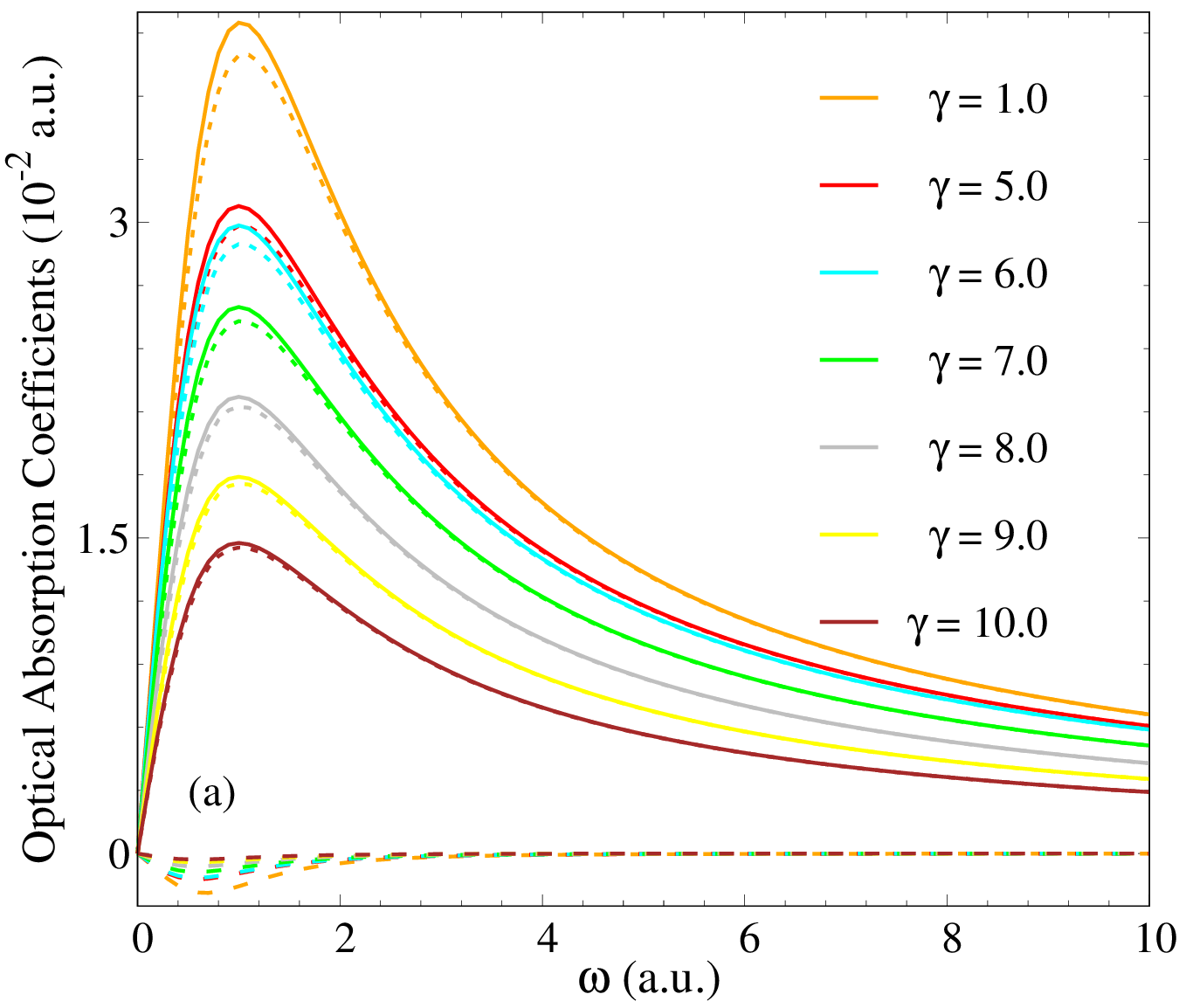}
} 
\quad 
{
\includegraphics[width=.4\textwidth]{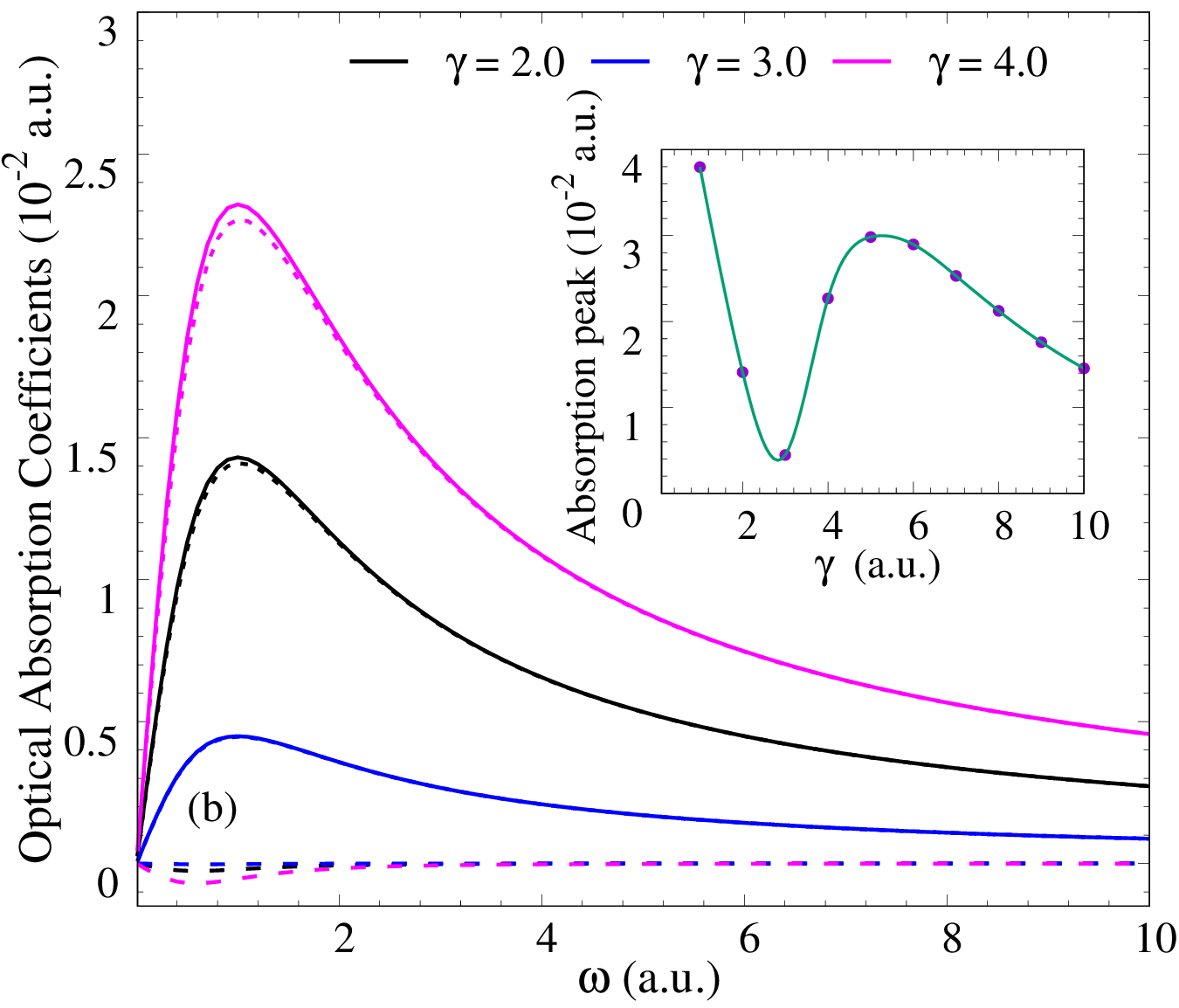}
} 
\caption{The linear (solid line), third-order non-linear (dashed line), and total absorption coefficient (dotted line) as a function of the photon energy for different values of the parameter $\gamma$, and $\omega_{0}=1.0$ of the quantum antidot potential. The inset stands for the peaks of the absorption coefficient as a function of the parameter $\gamma$.}
\label{fig:optical_g_case2}
\end{figure*}

\begin{figure*}[!t]
\centering
{
\includegraphics[width=.4\textwidth]{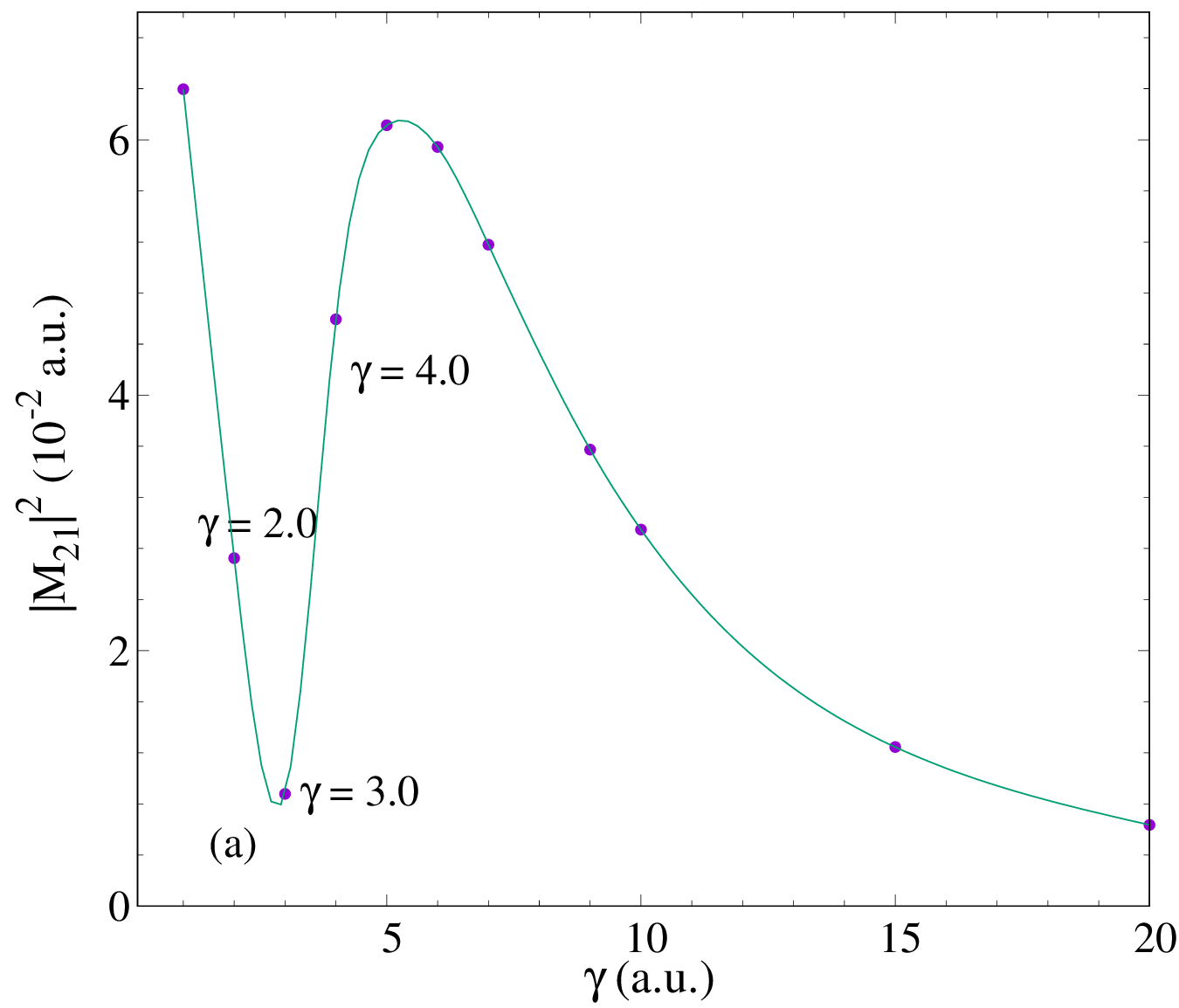}
} 
\quad 
{
\includegraphics[width=.4\textwidth]{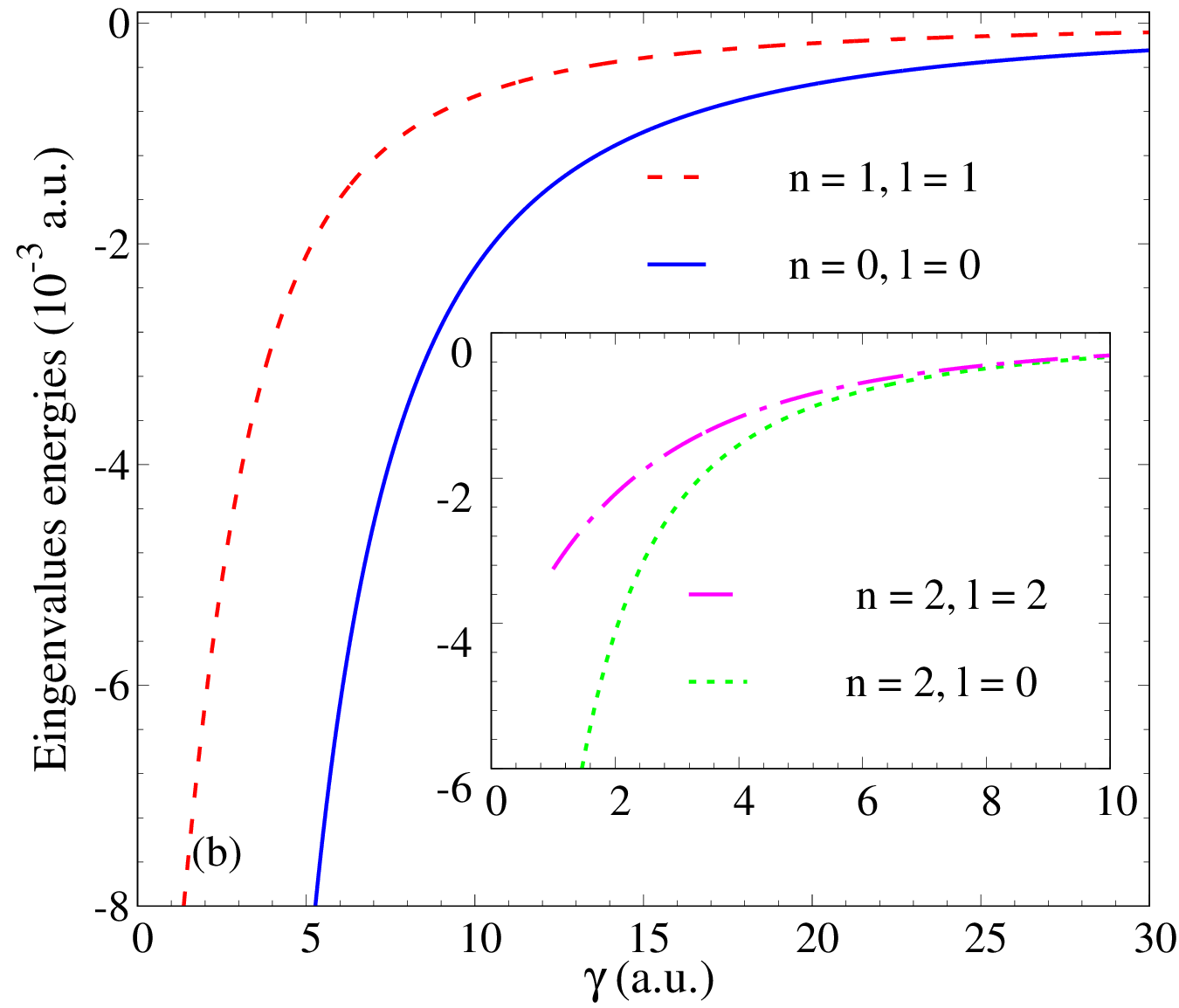}
} 
\caption{(a) Intrasubband dipole matrix elements of the dipole transition, and energy spectrum of the quantum antidot, as a function of the parameter $\gamma$, $\omega_{0}=1.0$, and $I=0.5$. The inset stands for the excited states.}
\label{fig:peak_dipole_g_case2}
\end{figure*}

\begin{figure}[!t]
\centering
\includegraphics[scale=0.4]{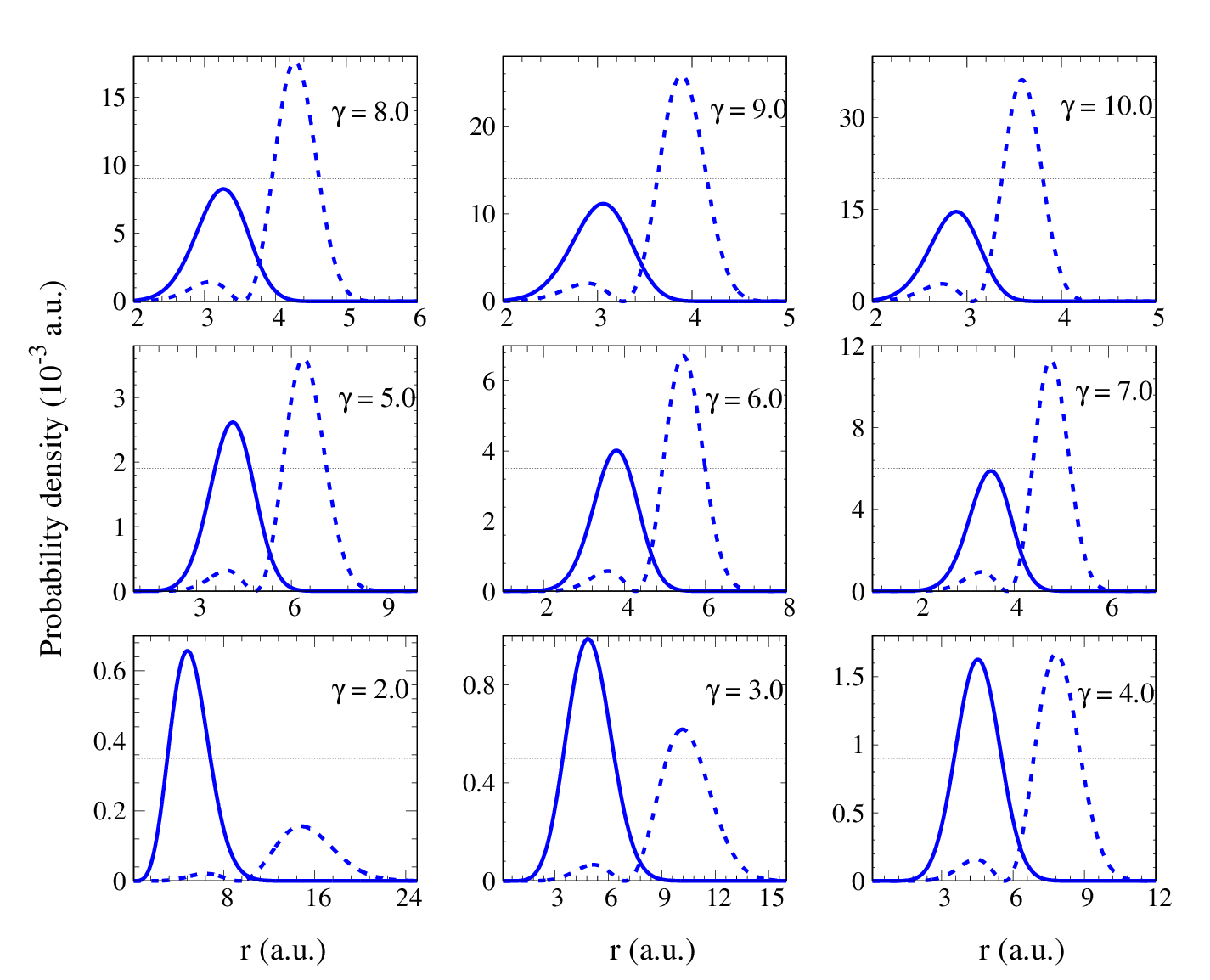}
\caption{The probability density of the quantum antidot states participating in the transition: solid line for the ground state ($n=0,l=0$), and dashed-line for the first excited state ($n=1,l=1$), as a function of the parameter $\gamma$, and $\omega_{0}=1.0$.}
\label{fig:wavefunction_g_case2}
\end{figure}
\begin{figure}[!t]
\centering
\includegraphics[scale=0.3]{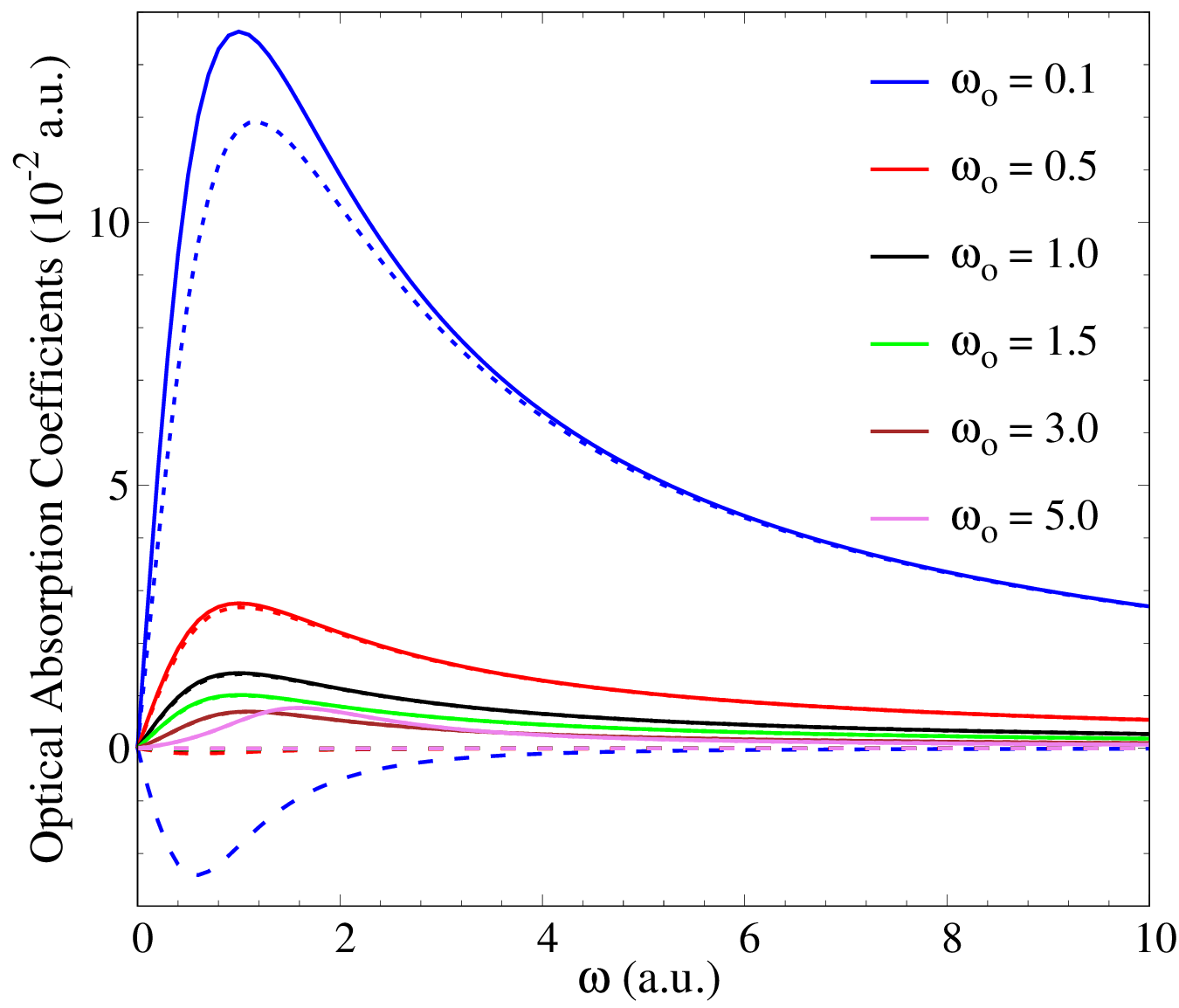}
\caption{Optical absorption coefficient of the quantum antidot as a function of the incident energy, for $\gamma=2.0$, $I=0.5$, and different values of the parameter  $\omega_{0}$.}
\label{fig:optical_case2_c}
\end{figure}

\begin{figure*}[!t]
\centering
{
\includegraphics[width=.4\textwidth]{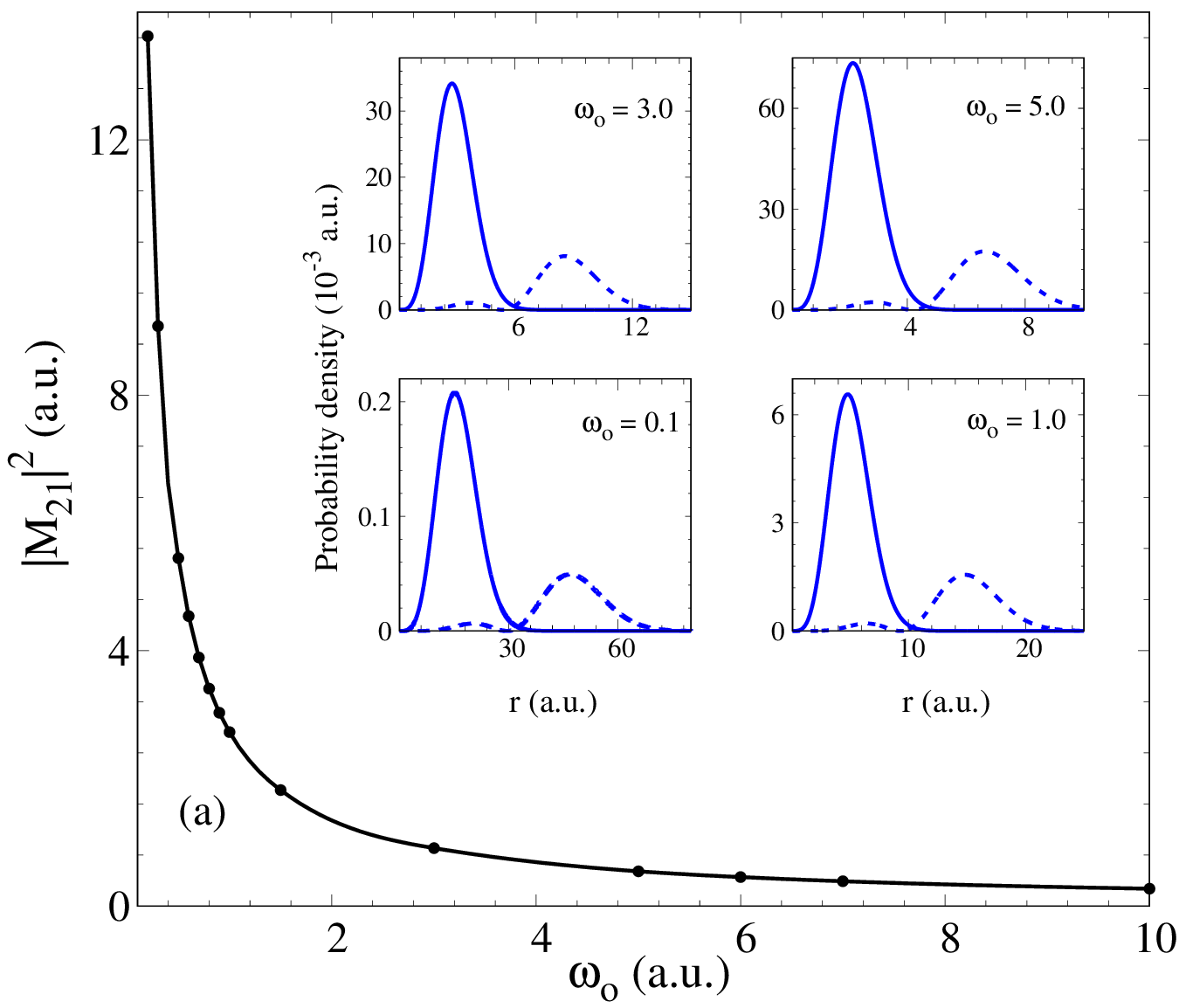}
} 
\quad 
{
\includegraphics[width=.4\textwidth]{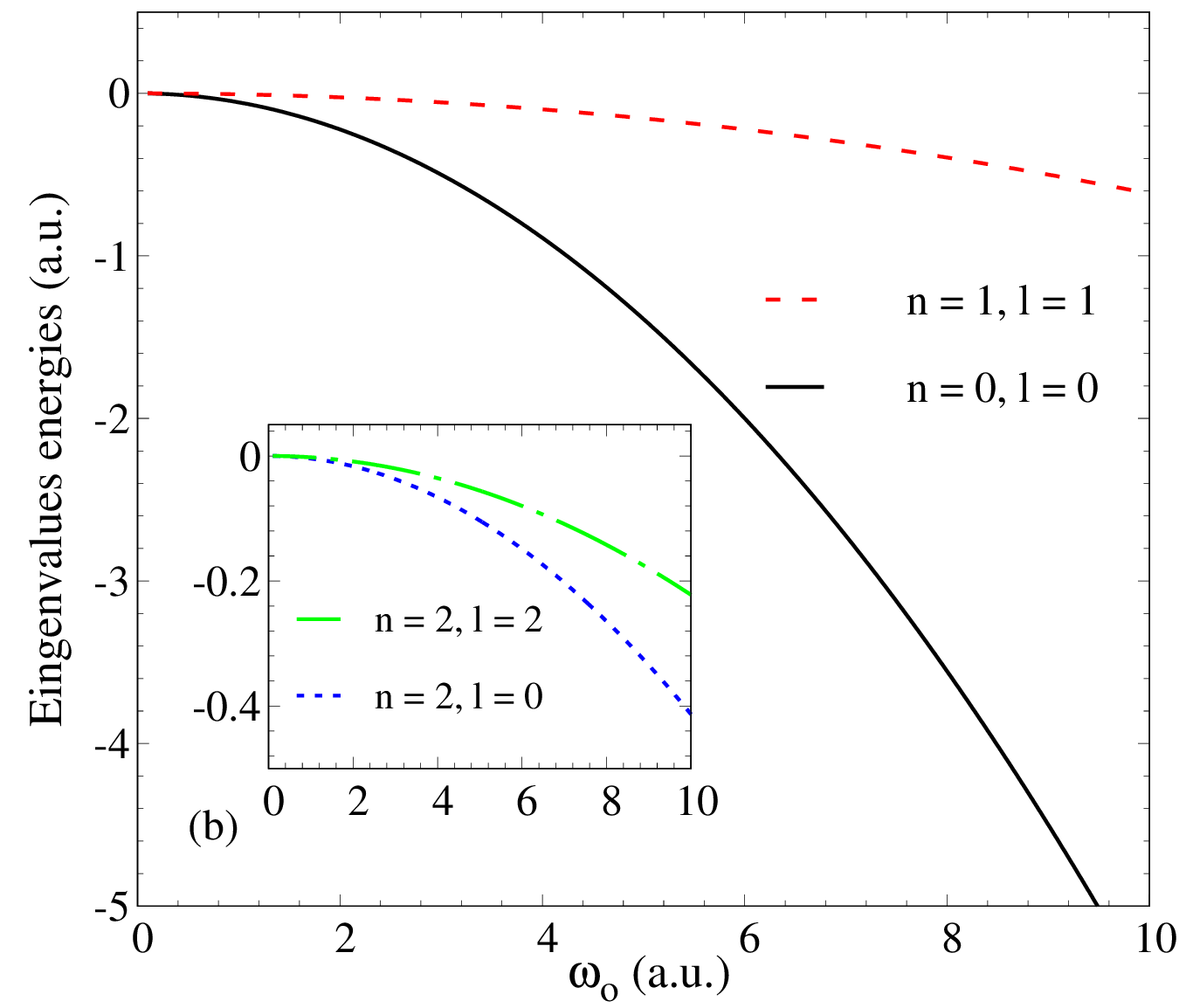}
} 
\caption{(a) Intrasubband dipole matrix elements of the dipole transition, and energy spectrum of the quantum antidot, as a function of the parameter $\omega_{0}$, $\gamma=2.0$, and $I=0.5$. The inset stands for (a) the wave function of the states participating in the transition and (b) the energies of the excited states.}
\label{fig:peak_dipole_energy_c_case2}
\end{figure*}

\begin{figure}[H]
\centering
\includegraphics[scale=0.3]{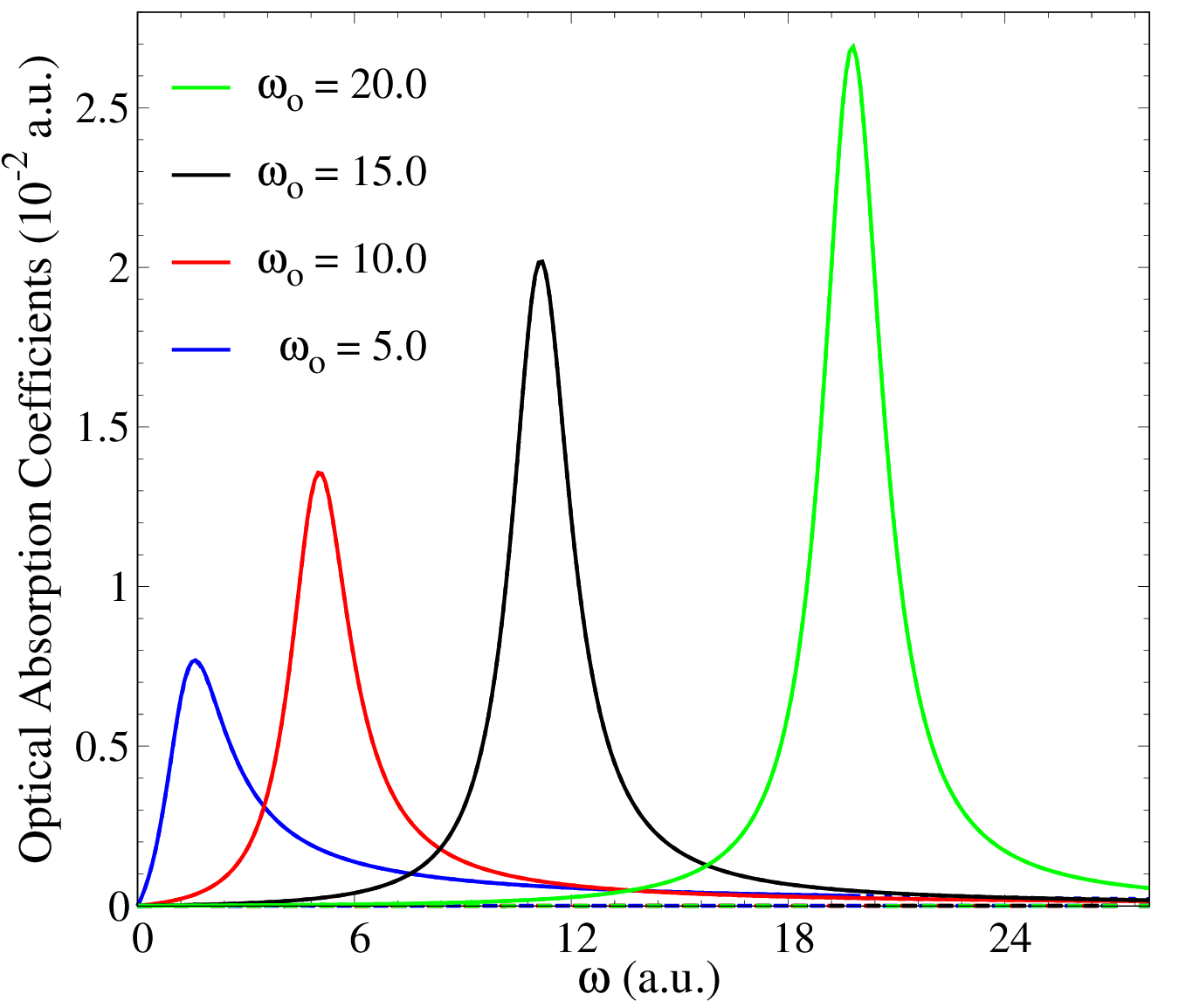}
\caption{The linear (solid line), third-order non-linear (dashed line), and total (dotted line) optical absorption coefficients of the quantum antidot, as a function of the photon energy for higher values of the parameter $\omega_{0}$, $\gamma=2.0$, and $I=0.5 \, (\text{a.u.})$.}
\label{fig:absorption_c_high_caso2}
\end{figure}

\begin{figure}[H]
\centering
\includegraphics[scale=0.3]{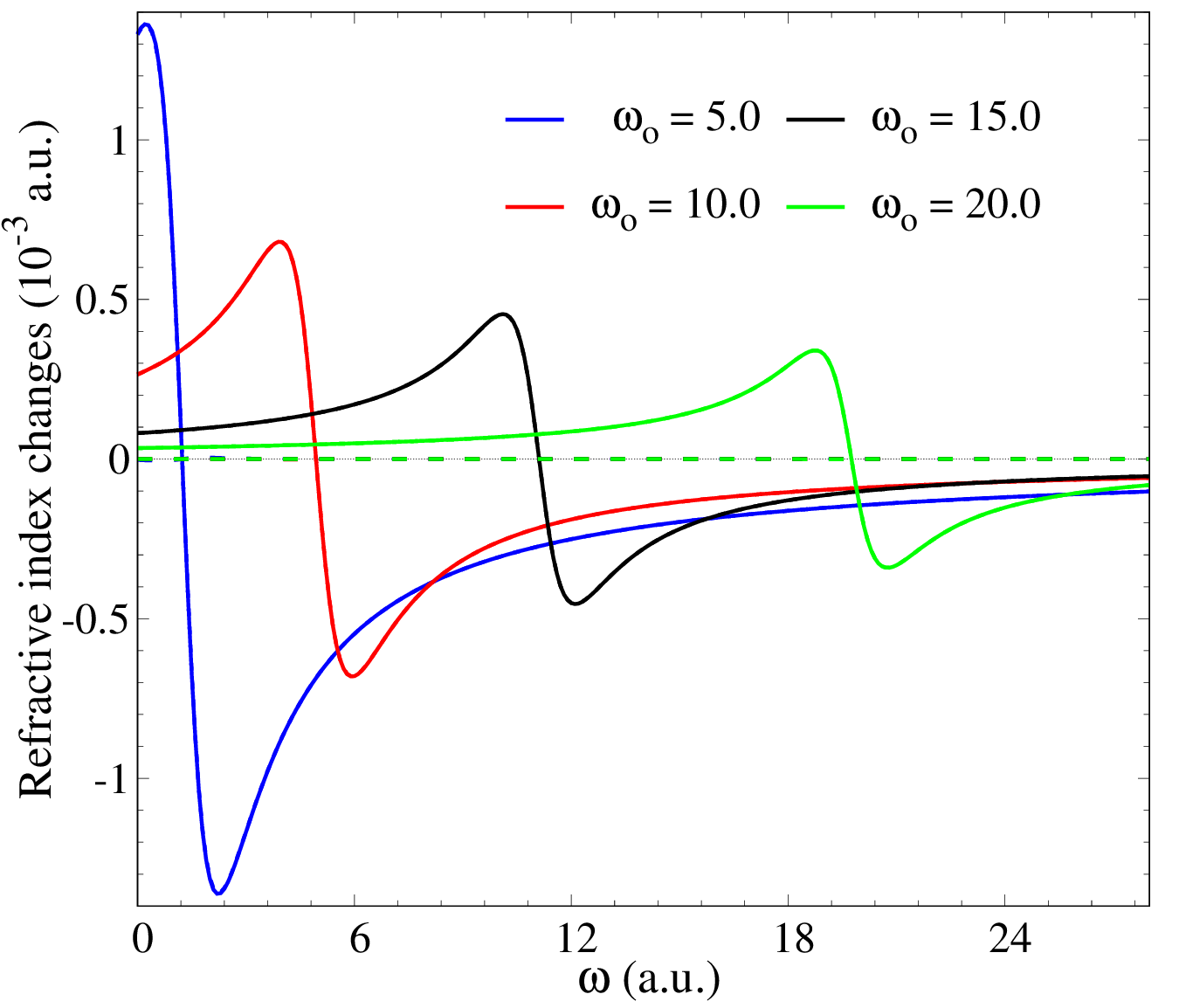}
\caption{The linear (solid line), third-order non-linear (dashed line), and total (dotted line) refractive index changes of the quantum antidot as a function of the photon energy for higher values of the parameter $\omega_{0}$, $\gamma=2.0$, and $I=0.5 \, (\text{a.u.})$.}
\label{fig:refractive_index_c_high_caso2}
\end{figure}

\subsection{\label{sec:case2_results}Case 2: quantum antidots}

We shall focus on the optical properties of quantum antidots modeled with a position-dependent mass. The system's behavior is analyzed under parameter variations $\gamma$ and the confining potential $\omega_0$, providing insights into the interplay between confinement and mass distribution. By studying the linear and nonlinear OAC and RIC, we aim to uncover how these properties evolve across different regimes of parameter space. In particular, we explore how the intrasubband dipole matrix elements and energy spectrum influence the system's optical response. Next, we discuss the effects of varying the parameters $\gamma$ and $\omega_0$ on the optical absorption coefficients and the refractive index changes in quantum antidots. Figure \ref{fig:optical_g_case2} shows the optical absorption coefficient, $\alpha^{(1)}(\omega)$, $\alpha^{(3)}(\omega, I)$, and $\alpha(\omega, I)$ as a function of incident photon energy $\hbar\omega$, for different values of parameter $\gamma$, and $\omega_{0}=1.0$. As we observed, increasing $\gamma$ decreases the absorption peak; however, its position remains unchanged, contrary to what was observed for the quantum dot. Note that the resonance absorption peak does not follow the same pattern for $\gamma$ values of $2.0,3.0$ and $4.0$. The inset could explain the physical reason in Fig. \ref{fig:optical_g_case2}(b), Fig. \ref{fig:peak_dipole_g_case2}(a), and Fig. \ref{fig:wavefunction_g_case2}: the peak of absorption is strongly influenced by the intersubband transitions, governed by the overlapping region (decreasing with increasing $\gamma$), and the extended region (increasing with decreasing $\gamma$) of the wavefunctions. Added to that, the intersubband energies rapidly decrease with increasing $\gamma$ (see Fig. \ref{fig:peak_dipole_g_case2}(b)).

In Figure \ref{fig:optical_case2_c}, $\alpha ^{(1)}(\omega)$, $\alpha ^{(3)}(\omega, I)$, and $\alpha(\omega, I)$ were calculated as a function of incident photon energy $\hbar\omega$ using different values of parameter $\omega_{0}$, and  $\gamma=2.0$.  As observed, the resonance peak of absorption decreases with the increase of the parameter $\omega_{0}$ while moving slightly to the right of the curve until it starts to increase when $\omega_{0}$ reaches nearly $5.0 \, \text{(a.u.)}$. After that, the resonance peak increases with increasing $\omega_{0}$ and moves towards higher photon energies, as illustrated in the Fig.\ref{fig:absorption_c_high_caso2}. Increasing $\omega_{0}$ increases the energy difference between the two subbands involved in the transition, as shown in Fig. \ref{fig:peak_dipole_energy_c_case2}. On the other hand, the RIC, Fig. \ref{fig:refractive_index_c_high_caso2} decreases linearly with the increasing $\omega_{0}$ and simultaneously shifts to the right.

\section{\label{sec:concl}Conclusions}

In this work, we employed a PCT method to investigate the influence of PDM on the optical properties of quantum dots and antidots. Through this approach, we derived the eigenvalue equations. We obtained the analytical expressions for both the energy levels and wavefunctions of the system, considering confining potentials that model quantum dots and antidots. The study was extended to compute the linear and nonlinear OAC and RIC, which were explored for different values of the PDM parameters.

Our results reveal that the optical properties of these systems are susceptible to variations in the position-dependent mass parameter $\gamma$. As $\gamma$ increases, the confinement effect becomes stronger, which shifts the optical absorption peaks towards higher energies while diminishing their magnitudes. Conversely, for small values of $\gamma$, the nonlinear contributions to the absorption and refractive index changes become more pronounced, leading to significant modifications in the optical response. These findings highlight the ability to fine-tune the optical behavior of quantum dots and antidots through careful control of the effective mass distribution.

The study provides critical insights into the interplay between confining potentials and position-dependent mass in low-dimensional systems, offering a pathway for designing optoelectronic devices with tailored optical properties. By manipulating the PDM parameters, one can optimize the efficiency of devices such as photodetectors, lasers, and other components in nanophotonics and quantum optics. This work opens avenues for further investigations into other material systems and more complex confinement potentials, where PDM effects could be exploited to engineer novel optical functionalities.

In summary, our exploration of the PDM framework advances the understanding of quantum dots and antidots and contributes to the broader effort to harness quantum mechanical effects in practical applications. The potential to adjust optical absorption and refractive index properties by modulating effective mass offers promising opportunities for next-generation optoelectronic device technologies.

\section*{\label{sec:acknw}Acknowledgement}

This work was partially supported by the Brazilian agencies CAPES, CNPq, FAPES, and FAPEMA. Edilberto O. Silva acknowledges the support from the grants CNPq/306308/2022-3, FAPEMA/UNIVERSAL-06395/22, FAPEMA/APP-12256/22. This study was partly financed by the Coordenação de Aperfeiçoamento de Pessoal de N\'{\i}vel Superior - Brazil (CAPES) - Code 001. C.  Filgueiras acknowledge FAPEMIG Grant No. APQ 02226/22. and  CNPq Grant No. 310723/2021-3.

\section*{\label{sec:author}Author contributions}
All authors contributed equally to the paper.

\section*{\label{sec:datar}Data Availability Statement}

Data will be made available on reasonable request.

\bibliographystyle{model1a-num-names}
\bibliography{mybibliography}

\end{document}